\documentclass[iop]{emulateapj}
\usepackage{apjfonts}
\usepackage{graphicx}
\usepackage{color}
\usepackage{amsmath}

\def\d3{$\delta_{3}$ }
\def\1d3{$(1 + \delta_{3})$ }
\def\l1d3{$\log_{10}(1 + \delta_{3})$ }

\def\s3{$\Sigma_{3}$}
\def\ha{H$\alpha$}
\def\hb{H$\beta$}

\def\24m{24 $\mu$m}

\def\sm{$\rm~M_{*}$}

\def\kms{${\rm km~s^{-1}}$ }

\def\Msolar{$\rm M_{\odot}$}

\def\fq{$F_{q}$}
\def\cq{$C_{q}$}
\def\rmxaa{RMxAA}

\def\sigsm{$\Sigma_{*}$}

\def\sigha{$\Sigma_{\rm H\alpha}$}
\def\h2{$\rm~H_{2}$}
\def\Mh2{$\rm~M_{H2}$}
\def\sigh2{$\Sigma_{\rm~H_{2}}$}

\def\finout{f$_{in-out}$}
\def\foutin{f$_{out-in}$}
\def\Re{$R_{e}$}

\shorttitle{Quenching}
\shortauthors{Lin et al.}

\begin{document}

\title{SDSS-IV MaNGA: Inside-out vs. outside-in quenching in different local environments}

\author{Lihwai Lin \altaffilmark{1}, Bau-Ching Hsieh \altaffilmark{1}, Hsi-An Pan \altaffilmark{1}, Sandro B. Rembold \altaffilmark{2,3}, Sebasti\'{a}n F. S\'{a}nchez \altaffilmark{4}, Maria Argudo-Fern\'{a}ndez \altaffilmark{5,6}, Kate Rowlands \altaffilmark{7}, Francesco Belfiore \altaffilmark{8},  Dmitry Bizyaev \altaffilmark{9,10}, Ivan Lacerna \altaffilmark{11,12}, Rogr\'{e}io Riffel \altaffilmark{13,3}, Yu Rong \altaffilmark{14,6,15},  Fangting Yuan \altaffilmark{16}, Niv Drory \altaffilmark{17}, Roberto Maiolino \altaffilmark{18,19}, Eric Wilcots \altaffilmark{20}}

\altaffiltext{1}{Institute of Astronomy \& Astrophysics, Academia Sinica, Taipei 10617, Taiwan; Email: lihwailin@asiaa.sinica.edu.tw}

\altaffiltext{2}{Departamento de F\'isica, CCNE, Universidade Federal de Santa Maria, 97105-900, Santa Maria, RS, Brazil}
\altaffiltext{3} {Laborat\'orio Interinstitucional de e-Astronomia, Rua General Jos\'e Cristino, 77 Vasco da Gama, Rio de Janeiro, Brazil, 20921-400}

\altaffiltext{4}{Instituto de Astronom\'ia, Universidad Nacional Aut\'onoma de  M\'exico, Circuito Exterior, Ciudad Universitaria, Ciudad de M\'exico 04510,  Mexico}

\altaffiltext{5}{Centro de Astronom\'ia, Universidad de Antofagasta, Avenida Angamos 601, Antofagasta 1270300, Chile}
\altaffiltext{6}{Chinese Academy of Sciences South America Center for Astronomy, China-Chile Joint Center for Astronomy, Camino El Observatorio, 1515, Las Condes, Santiago, Chile}

\altaffiltext{7} {Department of Physics \& Astronomy, Johns Hopkins University, Bloomberg centre, 3400 N. Charles St., Baltimore, MD 21218, USA}

\altaffiltext{8}{UCO/Lick Observatory, University of California, Santa Cruz, 1156 High St. Santa Cruz, CA 95064, USA}

\altaffiltext{9}{Apache Point Observatory and New Mexico State University, P.O. Box 59, Sunspot, NM, 88349-0059, USA}
\altaffiltext{10}{Sternberg Astronomical Institute, Moscow State University, Moscow, Russia}

\altaffiltext{11}{Instituto de Astronom\'ia, Universidad Cat\'olica del Norte, Av. Angamos 0610, Antofagasta, Chile}
\altaffiltext{12}{Instituto Milenio de Astrof\'isica, Av. Vicu\~na Mackenna 4860, Macul, Santiago, Chile}

\altaffiltext{13}{Departamento de Astronomia, Universidade Federal do Rio Grande
do Sul - Av. Bento Gon\c calves 9500, Porto Alegre, RS, Brazil}

\altaffiltext{14}{Instituto de Astrof\'isica, Pontificia Universidad Cat\'olica de Chile, Av. Vicu\~na Mackenna 4860, Macul, Santiago, Chile}
\altaffiltext{15}{Key Laboratory for Computational Astrophysics, National Astronomical Observatories, Chinese Academy of Sciences, 20A Datun Road, Chaoyang District, Beijing 100012, China}

\altaffiltext{16}{Shanghai Astronomical Observatory, Chinese Academy of Science, 80 Nandan Road, Shanghai 200030, China}

\altaffiltext{17}{McDonald Observatory, University of Texas at Austin, University Station, Austin, TX 78712-0259, USA}
\altaffiltext{18}{Cavendish Laboratory, University of Cambridge, 19 J. J. Thomson Avenue, Cambridge CB3 0HE, United Kingdom}
\altaffiltext{19}{University of Cambridge, Kavli Institute for Cosmology, Cambridge, CB3 0HE, UK.}
\altaffiltext{20}{Department of Astronomy, University of Wisconsin-Madison, 475N. Charter St., Madison WI 53703, USA}

\begin{abstract}
The large Integral Field Spectroscopy (IFS) surveys have allowed the classification of ionizing sources of emission lines on sub-kpc scales.  
In this work, we define two non-parametric parameters, quiescence (\fq) and its concentration (\cq), to quantify the strength and the spatial distribution of the quenched areas, respectively, traced by the  LI(N)ER regions with low EW(\ha). With these two measurements, we classify MaNGA galaxies into inside-out and outside-in quenching types according to their locations on the \fq~vs. \cq~plane and we measure the fraction of inside-out (outside-in) quenching galaxies as a function of halo mass. We find that the fraction of galaxies showing inside-out quenching increases with halo mass, irrespective of stellar mass or galaxy type (satellites vs. centrals). In addition, high stellar mass galaxies exhibit a greater fraction of inside-out quenching compared to low stellar mass ones in all environments. 
In contrast, the fraction of outside-in quenching does not depend on halo mass.
Our results suggest that morphological quenching may be responsible for the inside-out quenching seen in all environments. On the other hand, the flat dependence of the outside-in quenching on halo mass could be a mixed result of ram-pressure stripping and galaxy mergers. Nevertheless, at a given environment and stellar mass, the fraction of inside-out quenching is systematically greater than that of outside-in quenching, suggesting that inside-out quenching is the dominant quenching mode in all environments.

\end{abstract}


\section{INTRODUCTION}

It has been long recognized that environments play an important role in the galaxy evolution \citep{dre80,bal06,coo07,pen10,muz12,wet12}. Galaxies located in dense environments may experience short periods of mass assembly where different processes of star formation quenching are present such that their stellar populations are in general old with high metallicities as opposed to the field galaxies \citep[e.g.,][]{rob94,kau96,kun02,tra08,bla09}.

Studies of the main-sequence galaxies in various environments find that the specific star formation rate (sSFR)  of galaxies located in groups or clusters is systematically lower by 0.1 -- 0.3 dex when compared to that of the field galaxies \citep{vul10,hai13,lin14,jia17,jia18}. Nevertheless, recent works have concluded that the well-established `color-density' relation at low redshifts is primarily driven by the increase of the quiescent population, with the global reduction in the star formation rate of star-forming galaxies in dense environments only being a secondary effect \citep{bal04,koy13,lin14,lac16,jia17,arg18,lac18}. Combining all these results, it is suggested that there could be a mixing of various quenching processes that operate on different time-scales going on for group/cluster galaxies.

A number of environment-associated mechanisms have been proposed to explain the suppressed star formation rate in dense environments. For example, the gas in galaxy discs or halos can be stripped when galaxies fall into a cluster and move through the hot intracluster medium, often referred to as ram pressure stripping \citep{gun72,mc08}. This scenario is supported by the depleted HI gas and offset ionized gas observed in local cluster galaxies \citep[e.g.,][]{bro17,fri17}. Similarly, the outer hot halo of galaxies may also be
removed due to tidal forces or ram pressure, referred to as `strangulation', in which case galaxies lose their fuel to supply further star formation \citep{lar80}. Other processes, such as galaxy interactions or galaxy harassment \citep{moo96}, which are found more frequently in dense environments \citep{lin10}, may also induce gas inflow toward the centers of galaxies, trigger starbursts, and consume the gas completely \citep{mih94,cox06}. While it is likely that multiple processes may all contribute to the lower level of star formation activities  in dense environments, it is observationally very challenging to identify which mechanism is dominant over others. 

Recent integral field spectroscopy (IFS) surveys, such as CALIFA \citep{san12}, SAMI \citep{bry15}, and MaNGA \citep{bun15}, provide great opportunities to probe the quenching effect through spatially resolved information, as different quenching processes may leave distinct imprints on the spatial distributions of the star formation. For processes like ram-pressure
stripping or strangulation, one expects that the gas suppression and star formation quenching happen
outside-in or globally, while AGN feedback would result in the opposite trend (inside-out). With the resolved information, it is therefore possible to constrain the quenching mechanisms by investigating the spatial patterns of quenching within the galaxies \citep[e.g.,][]{gon14,gon16,tac15,li15,lin17b,ell18,san18}, including their dependence on the stellar mass, morphology, and environment \citep{per13,gon15,pan15,iba16,gon17,sch17,spi18,wan18,san18,med18}. Depending on the tracers used to study the quenching, some are more sensitive to the instantaneous halt of star formation (e.g., star formation rate), while others may in fact probe the ageing effect (e.g., age, D4000 strength, etc).

The environmental dependence of the resolved star formation activities so far have yielded controversial results. \citet{sch17} adopted the nearest neighbor local density as a environment tracer and studied the star formation rate gradients in the SAMI sample. They concluded that the star formation quenching occurs outside-in in dense environments. On the other hand, \citet{spi18} utilized galaxies taken from the MaNGA survey and found a global suppression in the star formation rate from inner to outer regions for satellite galaxies, which favors the strangulation scenario. 

In this paper, we investigate the spatial pattern of quenching and its dependence on the local environment, specifically the halo mass, in the MaNGA sample by quantifying the spatial distribution of quenched areas using non-parametric methods. We study the fractions of galaxies showing inside-out and outside-in quenching features as a function of halo mass for central and satellites galaxies separately, from which we infer the environment quenching mechanisms that act in massive halos.

Throughout this paper we adopt the following cosmology: \textit{H}$_0$ = 100$h$~\kms Mpc$^{-1}$, $\Omega_{\rm m} = 0.3$ and $\Omega_{\Lambda } = 0.7$. We use a Salpeter IMF and adopt the Hubble constant $h$ = 0.7. All magnitudes are given in the AB system.

\section{DATA \label{sec:data}}

\subsection{MaNGA IFU data}

MaNGA \citep{bun15,yan16b} is an integral field unit (IFU) survey on the SDSS 2.5m telescope (Gunn et al. 2006), as part of the SDSS-IV survey \citep{alb17,bla17}. MaNGA makes use of a modification of the BOSS spectrographs (Smee et al. 2013) to bundle fibres into hexagons (Drory et al. 2015). Each spectra has a wavelength coverage of 3500-10,000\AA,  and instrumental resolution $\sim$60 kms$^{-1}$ . After dithering, MaNGA data have an effective spatial resolution of 2.5\arcsec (FWHM; Law et al. 2015), and data cubes are gridded with 0.5\arcsec spaxels. 
 
In this study, we use $\sim$4690 galaxies with $z < 0.15$ taken from the MaNGA MPL-6 version of the internal release. To eliminate the effect of inclination to our analysis, we only use galaxies with a major to minor axis ratio (b/a) greater than 0.4 (i.e., excluding high-inclination systems with $i > 68^\circ$). This selection results in 4273 galaxies in our sample. We make use of the Pipe3D pipeline \citep{san16a} to model the stellar continuum with the GSD156 library of simple stellar populations \citep[SSPs;][]{cid13} that comprises 156 templates covering 39 stellar ages (from 1Myr to 14.1Gyr), and 4 metallicities (Z/Z$\odot$=0.2, 0.4, 1, and 1.5), extracted from  a combination of the synthetic stellar spectra from the GRANADA library \citep{mar05} and the MILES project \citep{san06,vaz10,fal11}. Details of the fitting procedures are described in \citet{san16b}. In short, a spatial binning is first performed in order to reach a S/N of
50 accross the entire field of view (FoV) for each datacube. A stellar
population fit of the coadded spectra within each spatial bin is then
computed. 
The stellar population model for spaxels with continuum S/N $>$ 3 is then estimated by
re-scaling the best fitted model within each spatial bin to the continuum
flux intensity in the corresponding spaxel, following \citet{cid13} and \citet{san16a}.  The stellar mass surface density (\sigsm) is obtained using the stellar mass derived for each spaxel and normalized to the physical area of one spaxel. The
best-fit stellar continuum is then subtracted from the reduced
data spectrum for the emission line measurements, which
are measured spaxel by spaxel using a weighted momentum analysis as described in \citep{san16b,san18}. All the emission lines were dust extinction corrected by using the Balmer decrement computed at each spaxel of the IFU cube, following the method described in the Appendix of \citet{vog13}. An extinction law with Rv = 4.5 \citep{fis05} and \citet{cal01} attenuation curve is used. These emission line measurements are later used for the ionizing source classification through the Baldwin-Phillips-Terlevich (BPT) 
excitation diagnostic diagrams \citep{bal81,vei87,kau03,kew06}. The classification is not sensitive to the dust reddening law since the lines in each pair used for the line ratio calculations are close enough in the wavelength.

\subsection{Halo mass}

The halo masses of our sample are adopted from the group catalog kindly made available by Yang et al. (in private communication). This catalog updates the SDSS DR4 group catalog of \citet{yan07,yan08} to the SDSS DR7 version. The galaxy groups are identified using an adaptive halo-based group finder based on the NYU-VAGC Catalog \citep{bla05}. The match to Yang's group catalog results in a subsample of 2915 galaxies with halo mass measurements.

For each galaxy, the group catalog provides two estimates for its halo mass: (1) $M_L$, based on the ranking of the characteristic luminosity $L_{19.5}$, the total luminosity of all group members with $^{0.1}M_r - 5 log h <= 19.5$; (2) $M_s$, based on the ranking of the characteristic stellar mass $M_{stel}$, the total stellar mass of all group members with $^{0.1}M_r - 5 log h <= 19.5$. Detailed tests with mock catalogs have shown that the halo masses are estimated reliably, with a standard deviation of about 0.3 dex \citep{yan08}. The two halo mass estimates yield very similar results for our analyses. In this paper, we present the results based on the luminosity-based halo mass $M_L$.

\begin{figure}
\centering
\includegraphics[angle=0,width=0.4\textwidth]{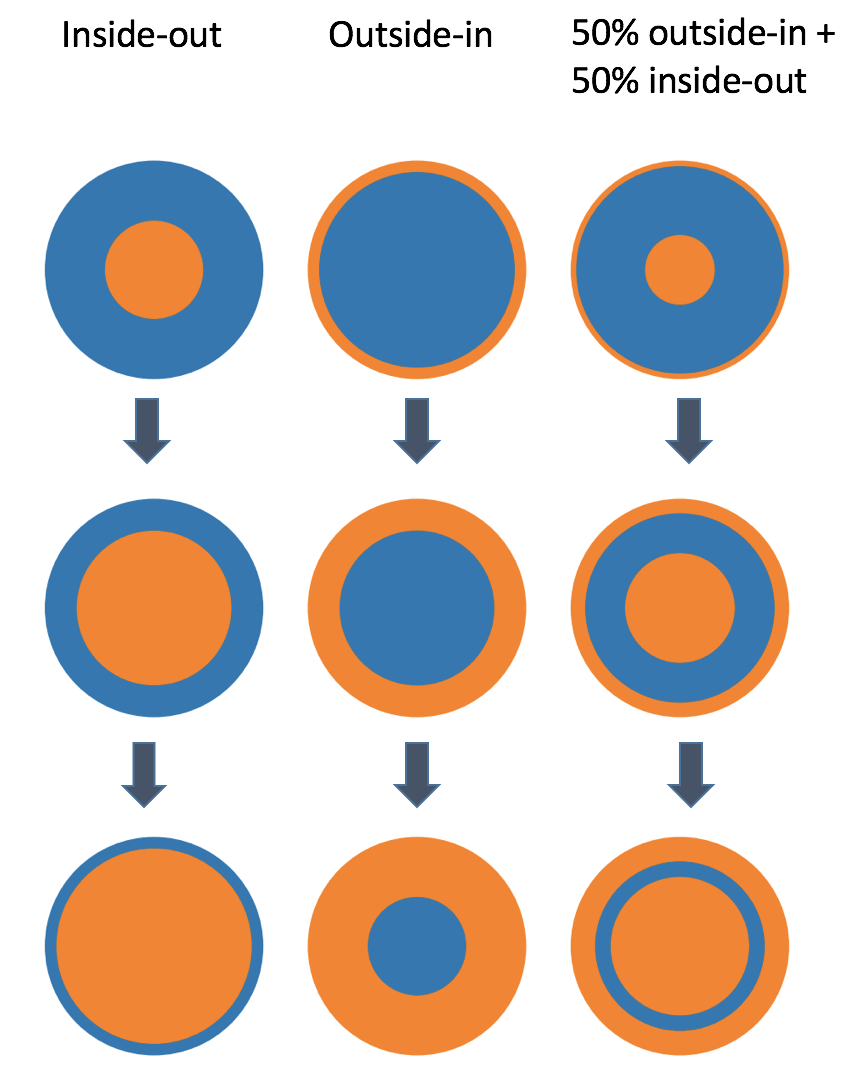}
\caption{ A schematic plot showing the time evolution of quenching used in our toy models (see Section 3 and Appendix A). Regions with ongoing star formation are shown in the blue color, whereas regions where the star formation has ceased is color-coded in red. Left sequence: inside-out quenching --star formation is quenched in the center first and then proceeds outwards. Middle sequence: outside-in quenching -- quenching proceeds inwards from the outer parts of galaxies. Right sequence: A mixture of 50\% inside-out quenching and 50\% outside-in quenching. \label{fig:model}}
\end{figure}

\begin{figure}
\centering
\includegraphics[angle=0,width=0.5\textwidth]{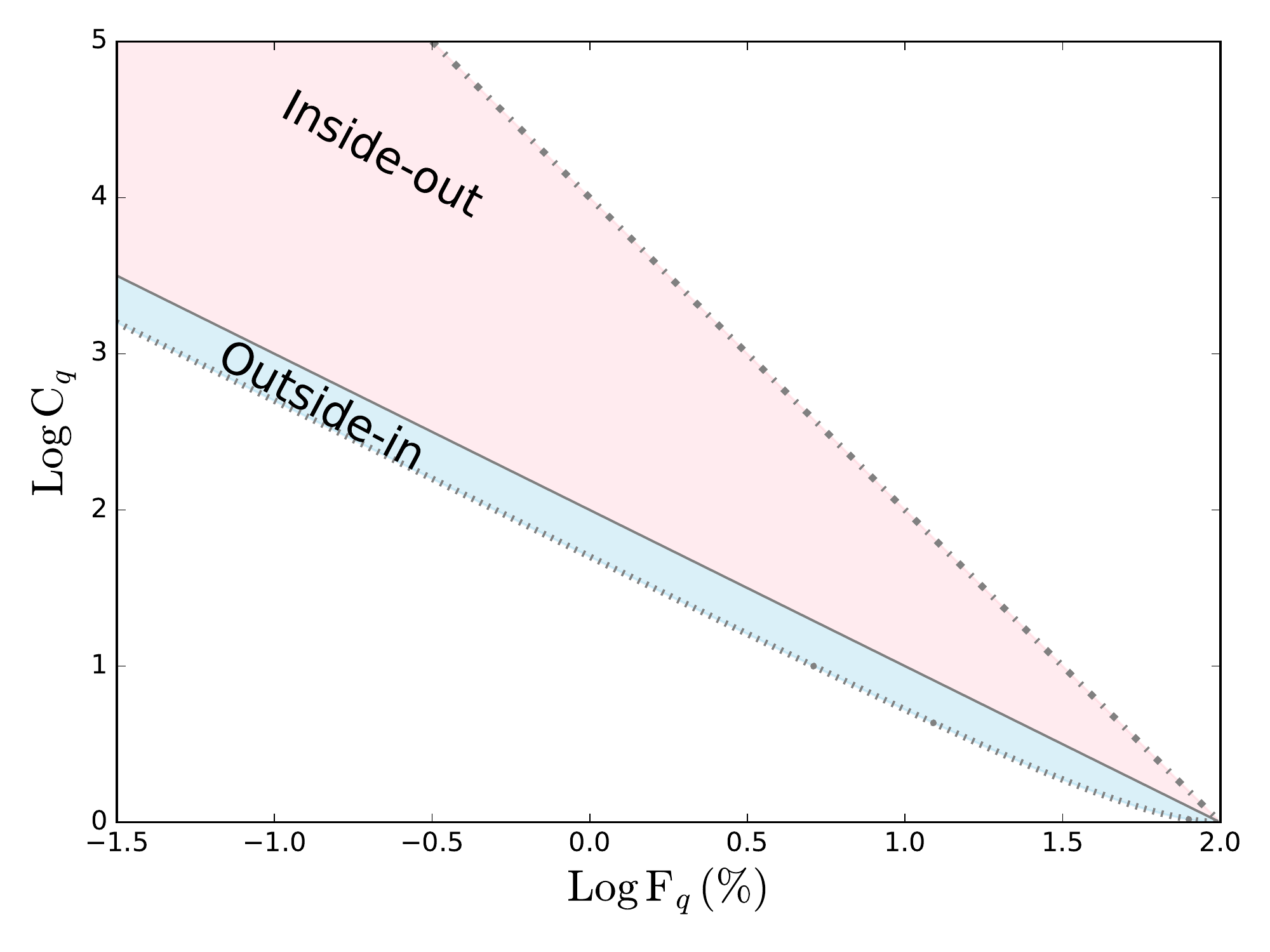}
\caption{The distributions of two toy models on the quiescence (\fq) vs. quenching concentration (\cq) plane. The dotted line represents the outside-in quenching sequence whereas the dotted-dashed line shows the inside-out quenching sequence. The solid line is the dividing line that separates the inside-out like (pink area) and outside-in like (light blue area) quenching modes (see the text of Section 4.2 and Appendex A) \footnote{As the toy models assume circular geometry and might be too simplistic, it is inevitable that some galaxies may fall outside the model boundaries, in which case the galaxies are still categorized according to whether they lie above or below the dividing line.}.   \label{fig:fq-cq_model}}
\end{figure}

\section{Methods}
With the advent of large IFU surveys, such as CALIFA, MaNGA, etc, ones are able to probe the emission line properties in sub-regions within the galaxies. 
In \citet{hsi17}, we confirmed the so-called `resolved star-forming main sequence' \citep{san13,can16,gon16}, the tight correlation between the star formation rate surface density and the stellar mass surface density (\sigsm) on kpc scales for HII regions classified based on the BPT line diagnostics. In addition, we found that the H$\alpha$ surface density (\sigha) is also strongly correlated with \sigsm~for regions classified as LI(N)ERs. The emission power of LI(N)ER regions is lower than that of the HII regions by nearly two orders of magnitude at a fixed \sigsm. The existence of this relation is in support of the scenario that LI(N)ER regions are primarily powered by hot evolved stars \citep[e.g.,][]{bin94,sta08,sar10,yan12,sin13,pap13,gom16,bel16,bel17}, whose abundance is proportional to the stellar mass. Since the emissions from hot evolved stars only begin to dominate after tens of Mys after the OB stars stop forming \citep{zha17} and the emission power of LI(N)ER is only a few percent of that of star-forming regions \citep{hsi17}, spaxels classified as LI(N)ER can be regarded as regions where the star formation has already ceased. This kind of association between the LI(N)ER regions and the `retired' or `quenched' areas has also been previously investigated in other works \citep{sin13,bel16,bel17}. Nevertheless, it is worth noting that some LI(N)ER-like emission may be associated with a weakly active galactic nucleus or shocks (e.g., in interacting galaxies), which normally have high equivalent widths (EW). In order to remove contributions from non-quenching origins of LI(N)ERs, we apply an EW (\ha) cut  when identifying the quenched areas (see the next paragraph).

For a given galaxy, we can then quantify the degree and the spatial distribution of the quenched areas by  looking into the abundance and locations of the quenched spaxels. In this work, we thus define two quantities, \fq~ and \cq, to describe the fraction and the concentration of the quenched areas within a galaxy, respectively. For each spaxel of MaNGA data, we first classify the emission line regions using the BPT diagrams.  We adopt the dividing curves suggested in the literature \citep[e.g.,][]{kew01,kau03,cid10} to separate various regions into HII, LI(N)ER, composite, and AGN regimes by using the [NII]/\ha~and [OIII]/\hb~ratio  as illustrated in the Figure 7 of \citet{lin17a} . To differentiate contributions between  quenched regions (or sometimes called `retired regions' ) and other ionizing sources (e.g., weakly AGNs or shocks) in powering LI(N)ERs, we further apply an EW(\ha) > -3\AA~ (positive value for absorption) cut in LI(N)ER spaxels when identifying final quenched areas \citep{cid11,hsi17}. Releasing the equivalent width criterion to  EW(\ha) > -6\AA~ \citep[e.g.,][]{san15} would result in an increase of inside-out quenching fraction (defined below) by $\sim 10$\% and change the outside-in quenching fraction with differences ranging from -50\% to +20\%. Nevertheless, this does not affect our main results and conclusions.

For each galaxy, we define the quenched fraction (or `quiescence' hereafter), \fq~, as the following:
\begin{equation}\label{eq:fq}
F_{q} = N_{quenched}/N_{all},
\end{equation}
where $N_{all}$ is the total number of spaxels within 1.5 $R_{e}$ with stellar mass surface density greater than $10^{6}$ \Msolar~ kpc$^{-2}$ and with at least one of the four emission lines (\ha, \hb, [NII], [OIII]) above the S/N threshold (3 for \ha~ and \hb; 2 for [NII] and [OIII]), and $N_{quenched}$ is the number of quenched spaxels (only galaxies with $N_{quenched} \geq 3$ are considered in this work).  
The stellar mass surface density cut is motivated by the data distribution on the star formation rate surface density vs. stellar mass surface density plane of \citet[][see their Figure 7]{abd18}. Our conclusions remain the same even if a higher cut of $10^{7}$ \Msolar~ kpc$^{-2}$ is applied.

The concentration of the quenched spaxels (hereafter `quenching concentration'), \cq, is computed as:
\begin{equation}\label{eq:cq}
C_{q} = \sum r_{all}^{2}/\sum r_{quenched}^{2},
\end{equation} 
where $r$ refers to the distance of a given spaxel to the galaxy center, corrected for the inclination. While \fq~represents the degree of quenching in a galaxy, \cq~reflects the spatial distribution of the quenched area. 

The combination of \fq~and \cq~ provides a powerful method for describing the spatial sequence of quenching. At a fixed \fq, galaxies with quenching occurring in the inner regions have a greater value of \cq~than those with quenching occurring in the outskirts. In other words, the inside-out quenching and outside-in quenching will follow different trajectories in the \fq~vs. \cq~diagram. To illustrate this point, we perform  two sets of toy models, one with inside-out quenching, starting from the inner spaxels to the outer spaxels (see the left sequence in Figure \ref{fig:model}) and the other with outside-in quenching, proceeding with the opposite direction (see the middle sequence in Figure \ref{fig:model}), assuming a perfect circle. For each set of the models, we create 80 equally-spaced annulus bins. In each step, we shift the annulus boundary that separates the quenched and unquenched areas by one annulus bin and compute \fq~ and \cq. The obtained inside-out and outside-in quenching trajectories are shown as the dot-dashed line and the dotted line in Figure \ref{fig:fq-cq_model}, respectively.  In fact, these two trajectories can also be derived analytically. In Appendix A, we consider a general case where there is a mixture of inside-out and outside-in quenching within a galaxy (see the right sequence in Figure \ref{fig:model}). The equations describing the pure inside-out and outside-in quenching lines can be obtained using Equation \ref{eq:math} by adopting $F_{qI}$ (the contribution of inside-out quenching) = 100 and 0, respectively:
\begin{align}
\log_{10} C_{q} & =  -2 \log_{10} F_{q} + 4    & \textsf{(Inside-out)}\label{eq:inout}\\
\log_{10} C_{q} & =  \log_{10} \frac{1}{1-\left(1-F_{q}/100\right)^2} & \textsf{(Outside-in)}\label{eq:outin}
\end{align}

By comparing the locations of the observed \fq~and \cq~of galaxies with the models, we are able to categorize whether a galaxy is more inside-out quenching like or outside-in quenching like. The criteria we adopted for the selection of quenched areas only concern the stellar mass surface density, the line equivalent width, and the line ratios, all of which do not directly depend on the observation resolution. As the two parameters (\fq~and \cq) deal with the relative quantities, our method is not that sensitive to the spatial resolution in the case where there are sufficient resolution elements.  Nevertheless, it is worth noting that very compact regions of quenched areas could be missed in our data and can only be resolved with greater spatial resolutions. Higher resolution observations would be required to investigate whether this is an important effect or not.

\section{Results} 

\begin{figure}
\centering
\includegraphics[angle=0,width=0.5\textwidth]{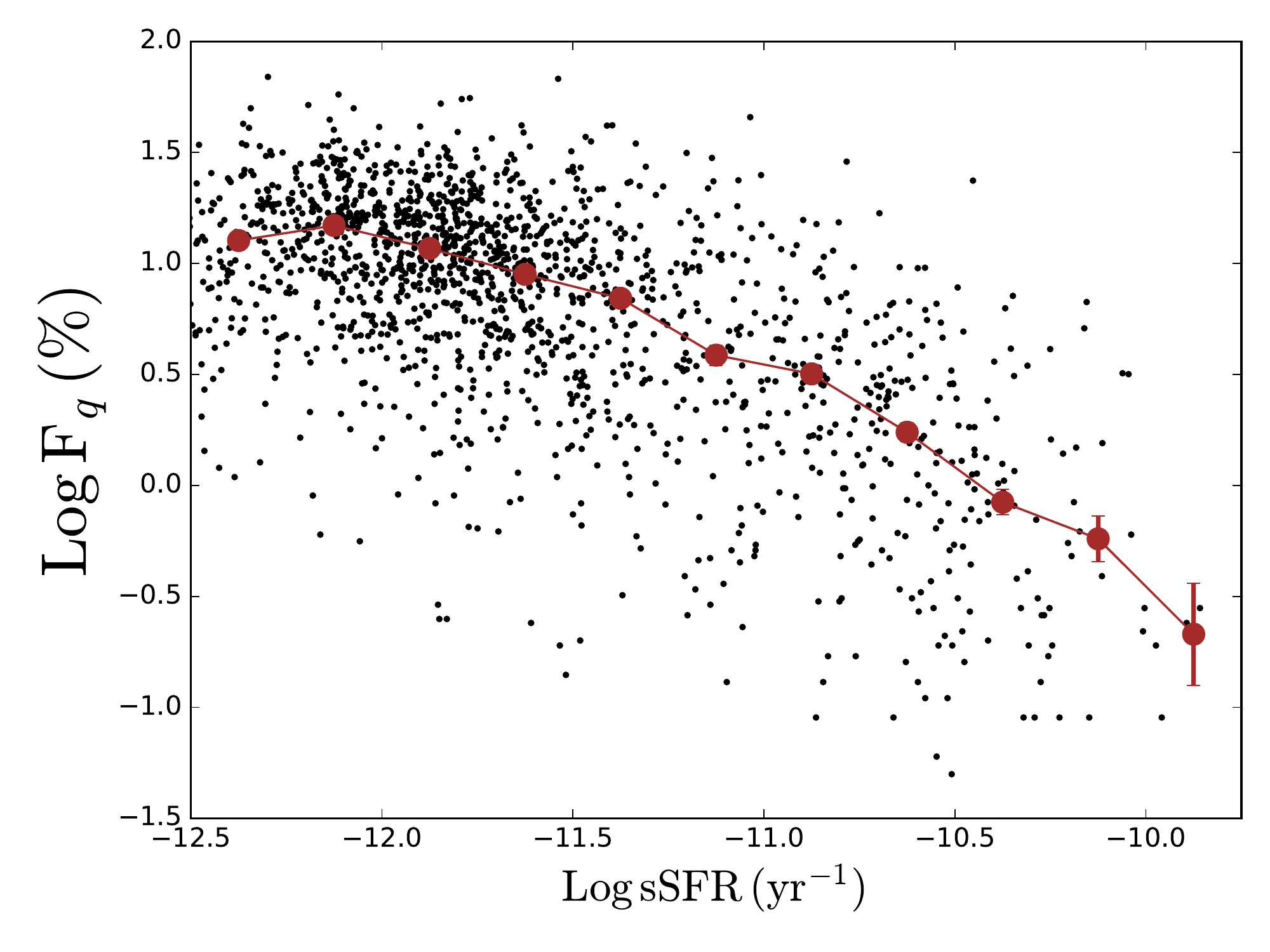}
\caption{Quiescence (\fq) vs. sSFR for the MaNGA sample. Black dots represent the measurements of each individual galaxy. The brown symbols show the medians and the associated uncertainties (computed as the root-mean-square in the logrithm space normalized by the square root of the sample size in each bin) of galaxies with measurable \fq. \label{fig:fq-ssfr}}
\end{figure}

\begin{figure}
\centering
\includegraphics[angle=0,width=0.5\textwidth]{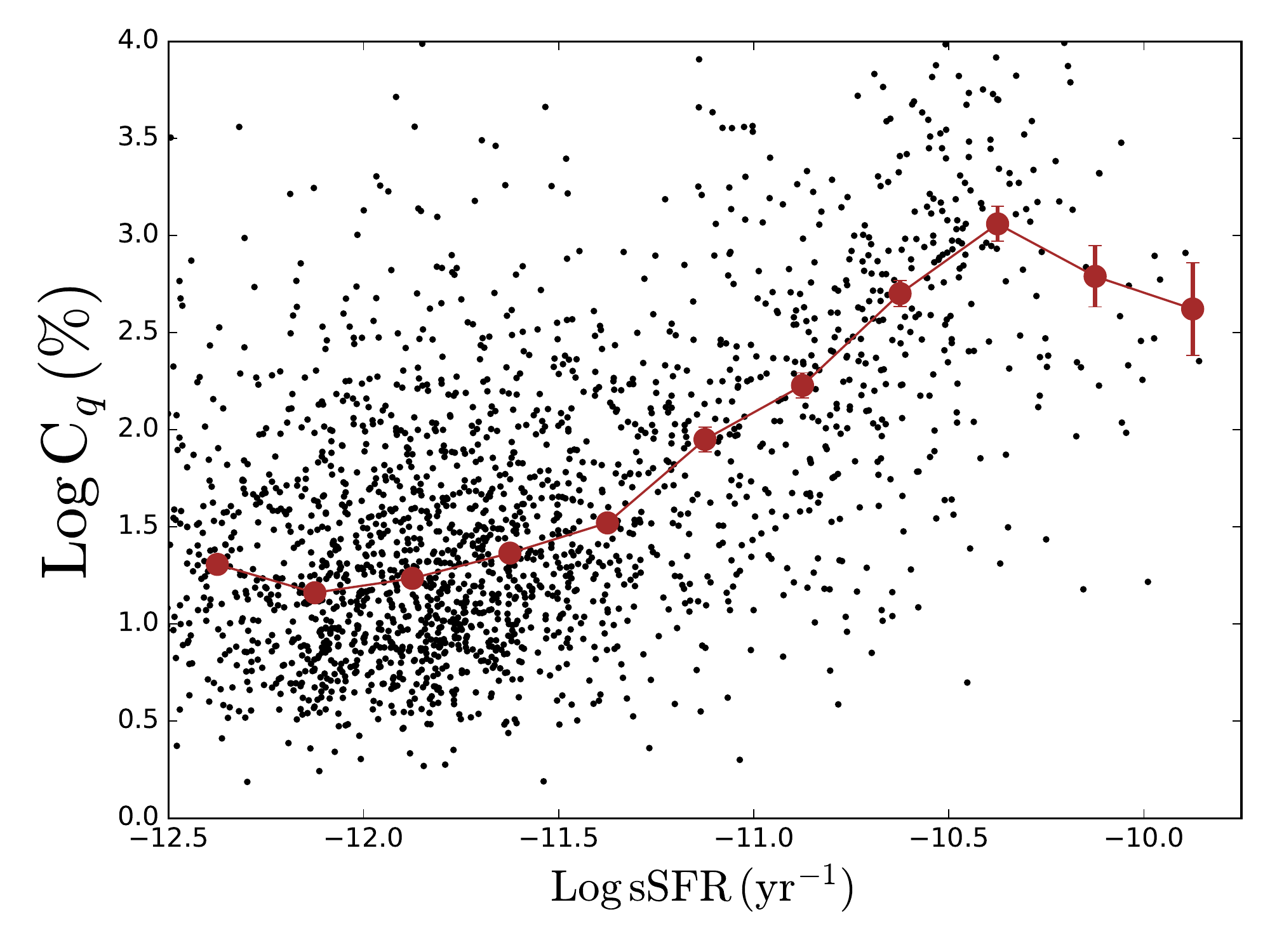}
\caption{Concentration of quenched area (\cq) vs. sSFR for the MaNGA sample. Black dots represent the measurements of each individual galaxy. The brown symbols show the medians and the associated uncertainties (computed as the root-mean-square in the logrithm space normalized by the square root of the sample size in each bin) of galaxies with measureable \cq. \label{fig:cq-ssfr}}
\end{figure}

\begin{figure}
\centering
\includegraphics[angle=0,width=0.5\textwidth]{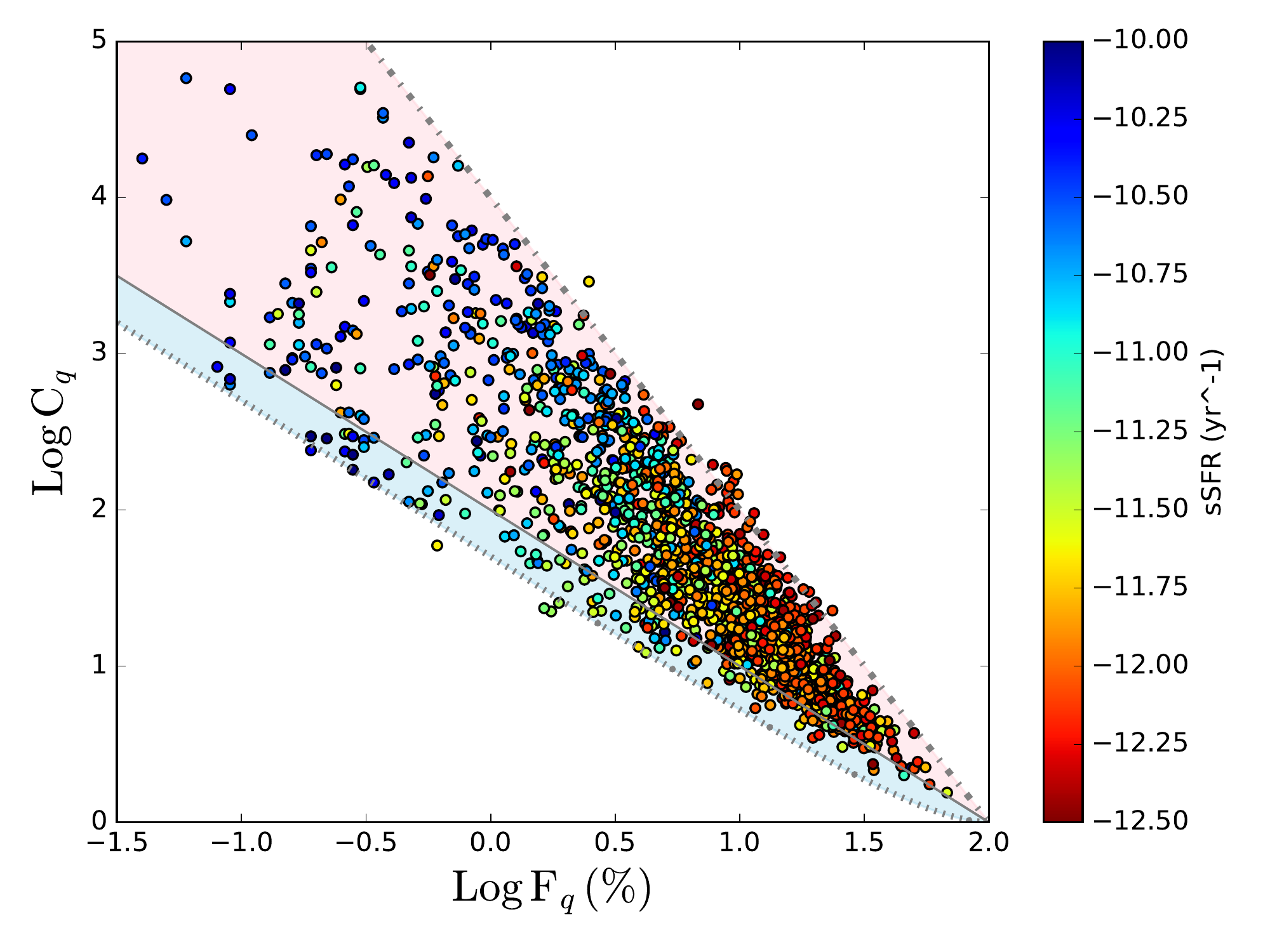}
\caption{The distributions of MaNGA galaxies on the quiescence (\fq) vs. quenching concentration (\cq) plane, color-coded according to the their sSFR. Similar to Figure \ref{fig:fq-cq_model}, the light blue shaded area denotes the outside-in like quenching mode whereas the pink shaded region denotes the inside-out like quenching mode.   \label{fig:fq-cq_ssfr}}
\end{figure}

\begin{figure}
\centering
\includegraphics[angle=0,width=0.5\textwidth]{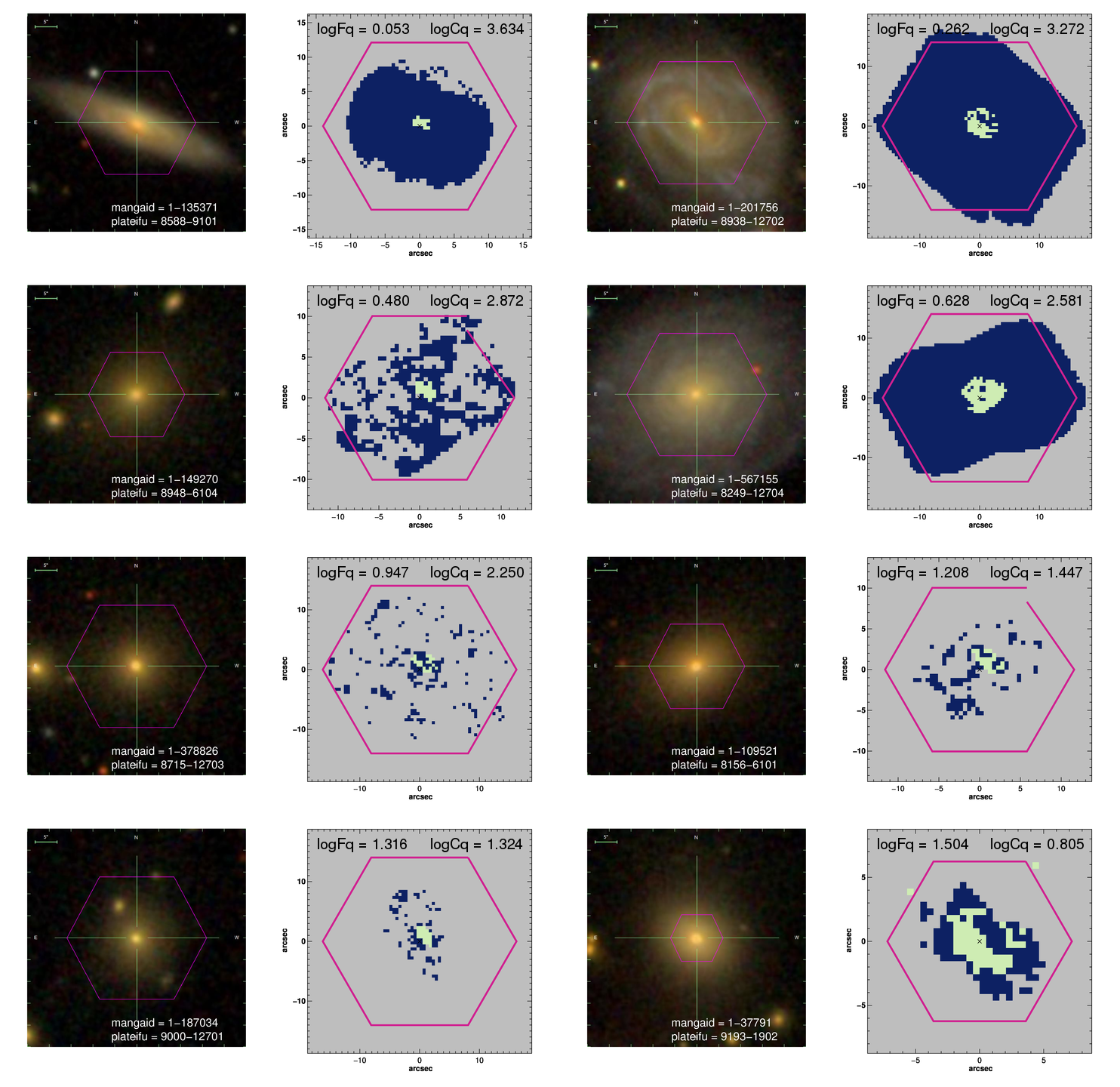}
\caption{Examples of galaxies classified as inside-out-like quenching. The first and third columns show the SDSS $gri$ composite images. The spatial distributions of the quenched areas are shown in yellow in the second and fourth columns, associated with the objects in the first and third columns, respectively. \label{fig:img_inout}}
\end{figure}

\begin{figure}
\centering
\includegraphics[angle=0,width=0.5\textwidth]{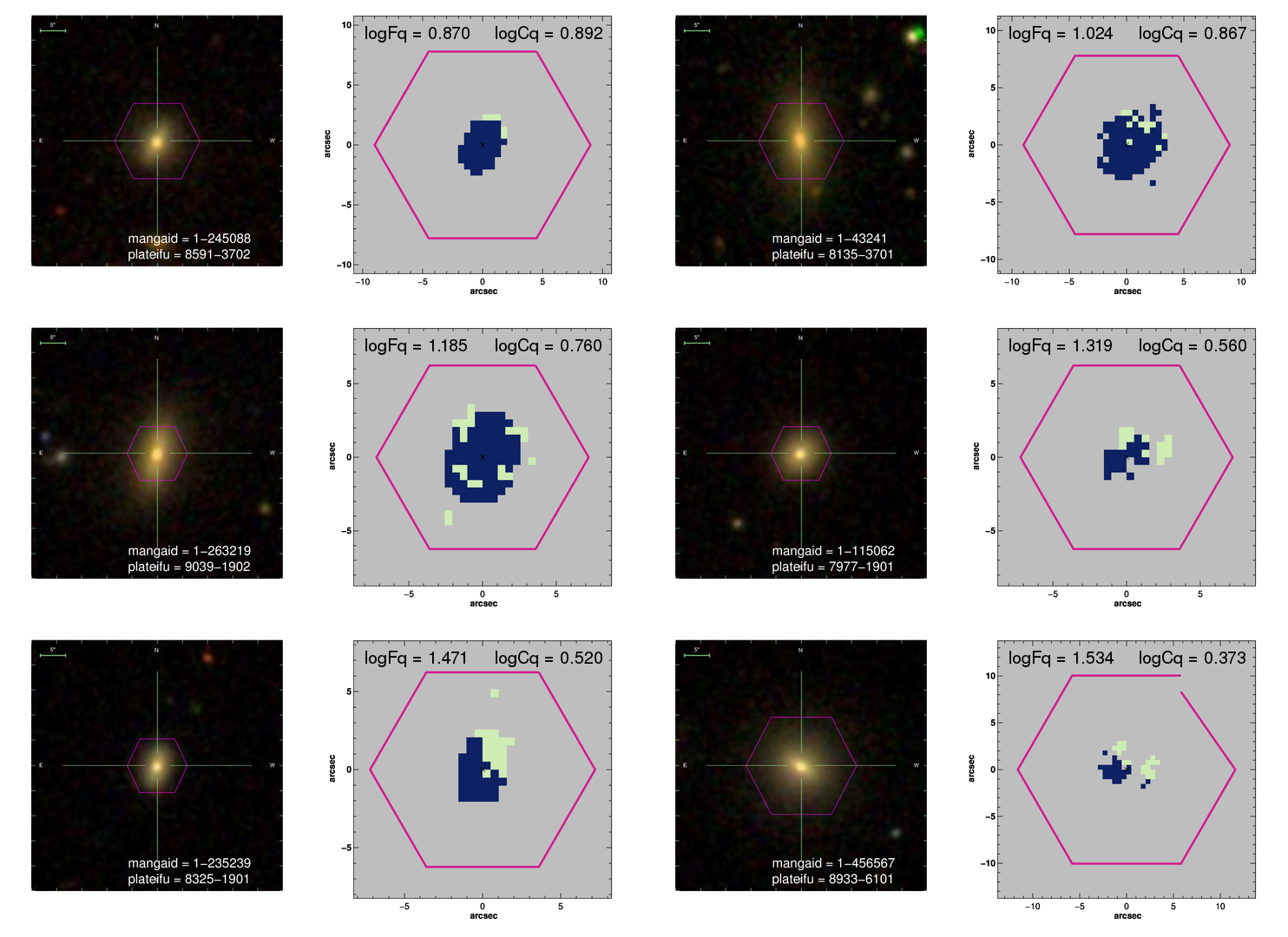}
\caption{Examples of galaxies classified as outside-in-like quenching. The first and third columns show the SDSS $gri$ composite images. The spatial distributions of the quenched areas are shown in yellow in the second and fourth columns, associated with the objects in the first and third columns, respectively. \label{fig:img_outin}}
\end{figure}

\begin{figure}
\centering
\includegraphics[angle=0,width=0.5\textwidth]{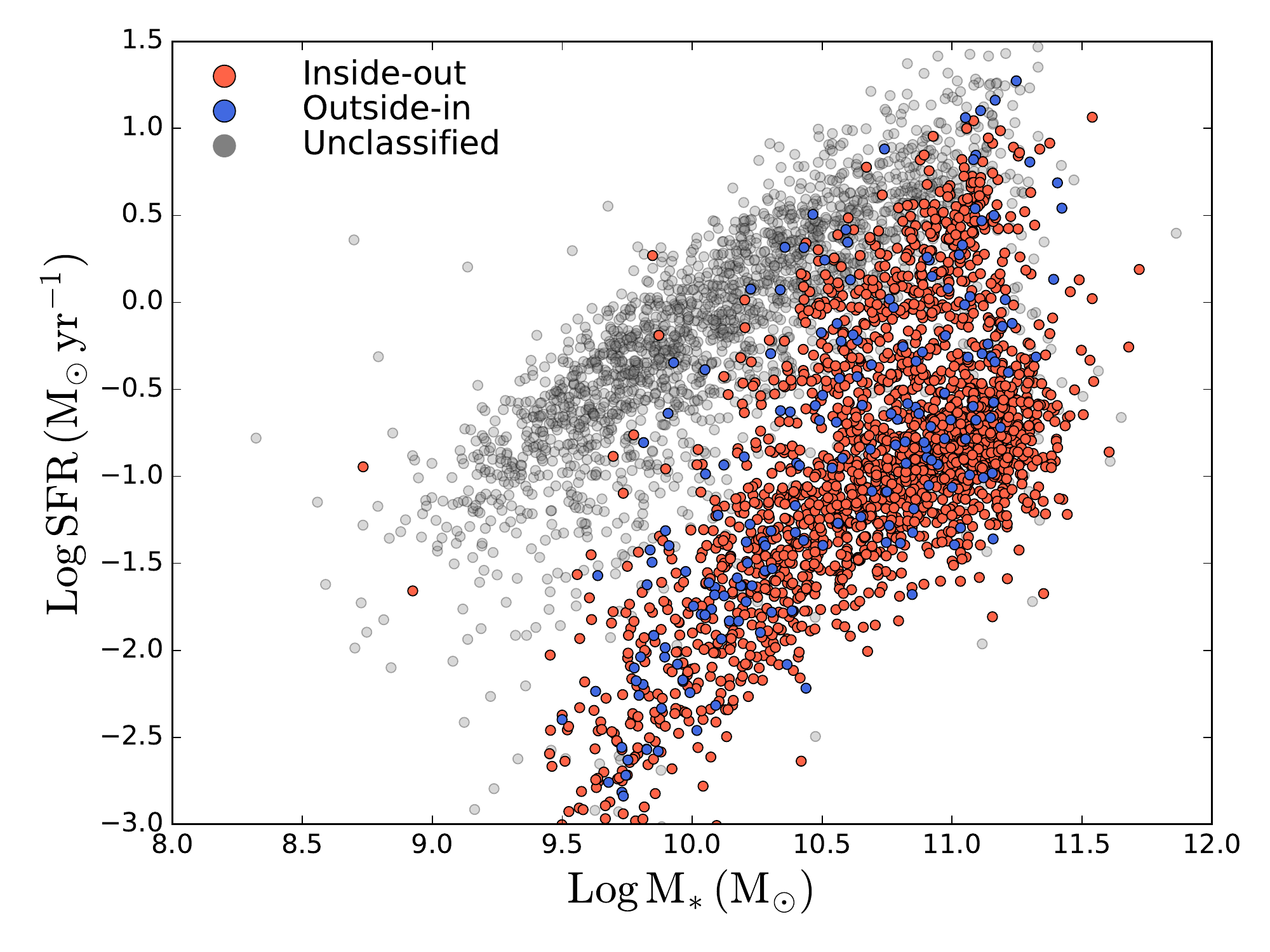}
\caption{The distributions of MaNGA galaxies on the SFR vs. \sm~ plane. The red and blue symbols denote galaxies classified as inside-out and outside-in quenching like objects, respectively. Unclassified galaxies are shown as grey points.   \label{fig:sm-sfr_quench}}
\end{figure}

\begin{figure*}
\centering
\includegraphics[angle=0,width=0.95\textwidth]{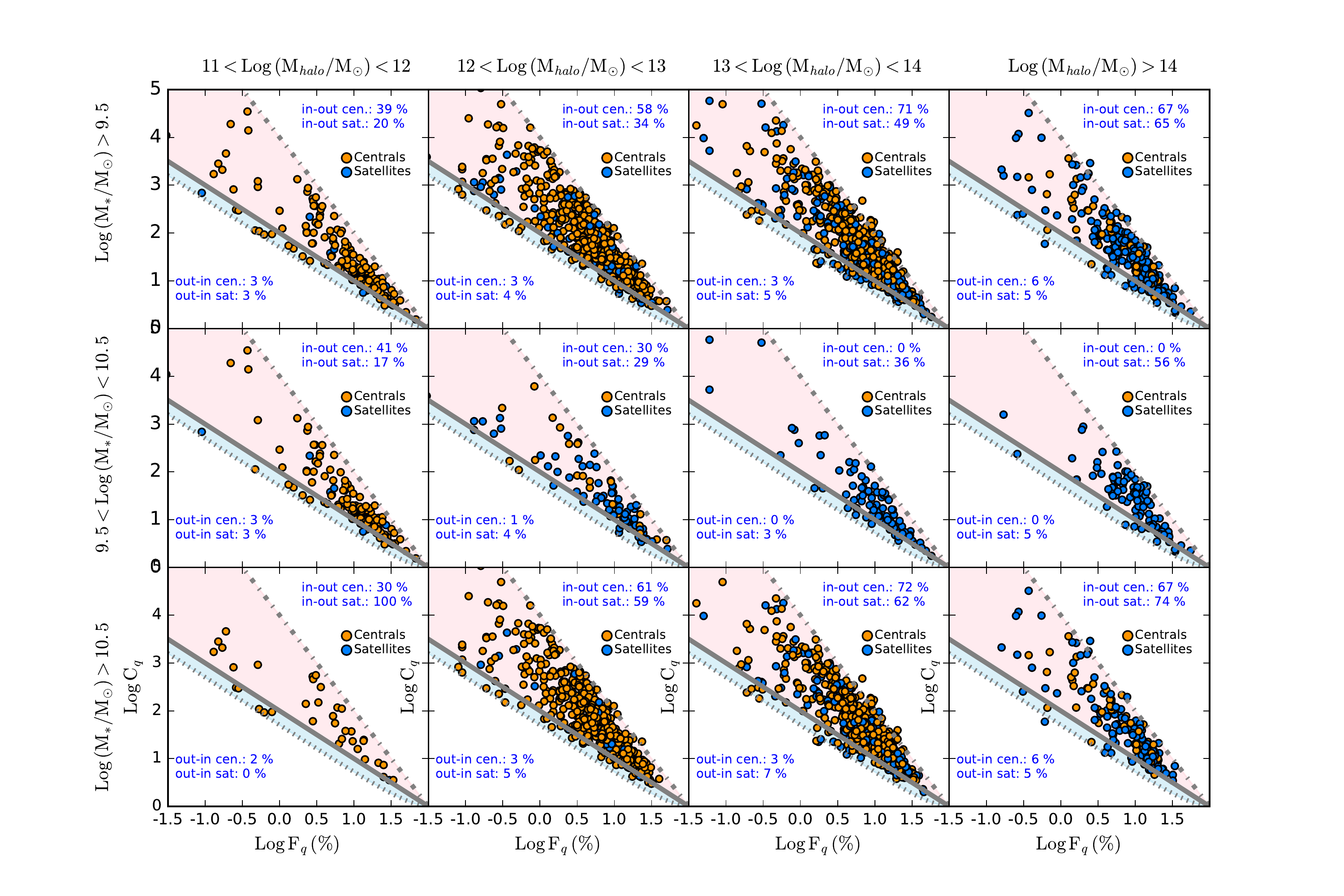}
\caption{The distributions of all MaNGA galaxies on the quiescence (\fq) vs. quenching concentration (\cq) plane in different halo mass bins (increasing from left to right). The centrals and satellites are shown in orange and blue symbols, respectively. Similar to Figure \ref{fig:fq-cq_model}, the light blue shaded area denotes the outside-in like quenching mode whereas the pink shaded region denotes the inside-out like quenching mode. Top, middle, and bottom panels are for galaxies with different stellar mass cuts (from top to bottom:  all galaxies with \sm~$> 10^{9.5}$\Msolar, $10^{9.5}$\Msolar~ $<$ \sm~ $< 10^{10.5}$\Msolar, and \sm~$> 10^{10.5}$\Msolar)
\label{fig:fq-cq_halo}}
\end{figure*}

\begin{figure*}
\centering
\includegraphics[angle=0,width=0.9\textwidth]{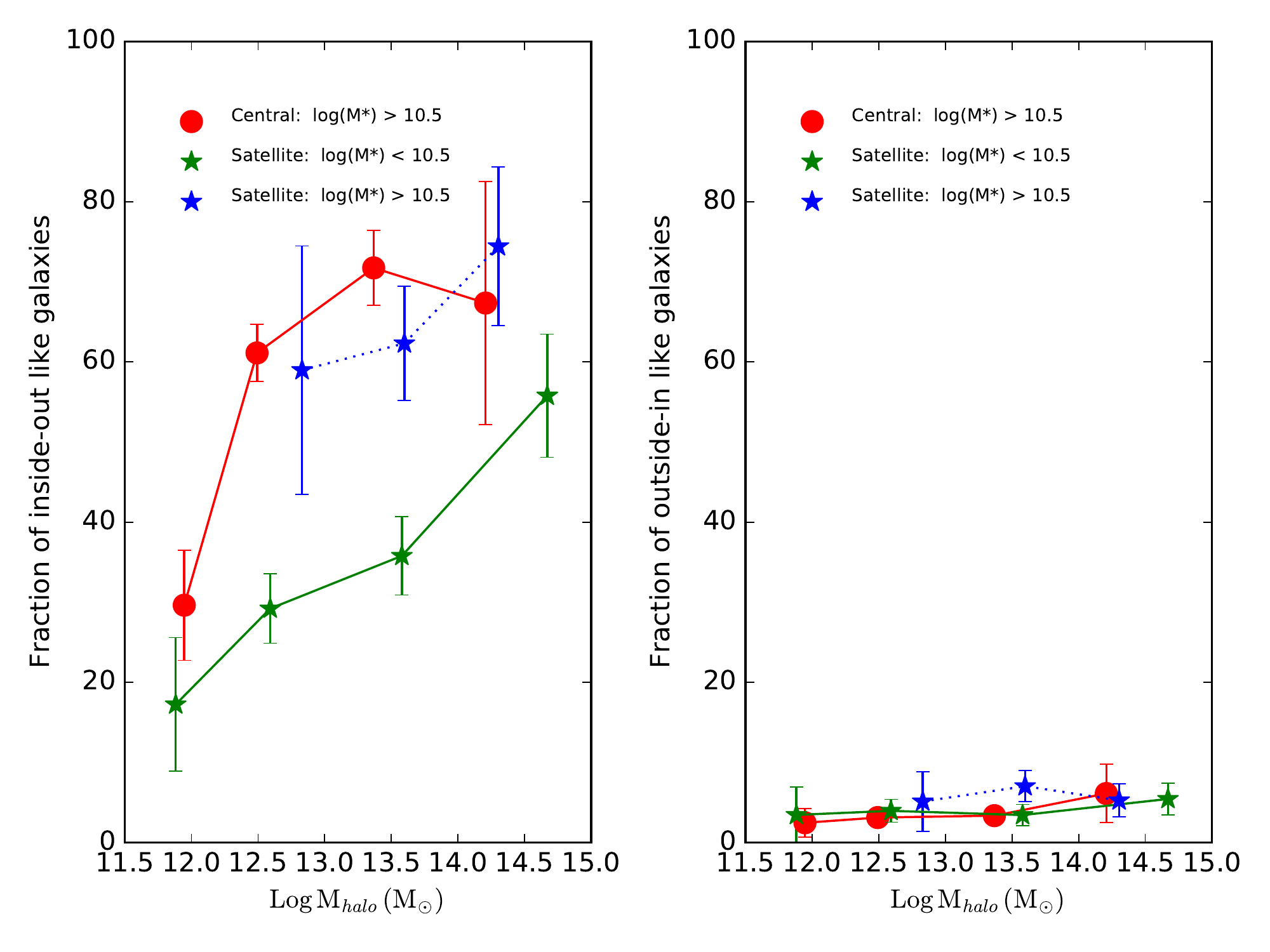}
\caption{The fraction of inside-out like (left panel) and outside-in like (right panel) galaxies of centrals with \sm~$> 10^{10.5}$ \Msolar (red symbols), satellites with \sm~$< 10^{10.5}$ \Msolar~ (green symbols), and satellites with \sm~$> 10^{10.5}$ \Msolar~ (blue symbols). The dotted and solid lines correspond to high-mass and low-mass galaxies, respectively. The error bars are computed as the root-mean-square normalized by the square root of the sample size in each bin. \label{fig:frac-halo}}
\end{figure*}

\begin{figure}
\centering
\includegraphics[angle=0,width=0.5\textwidth]{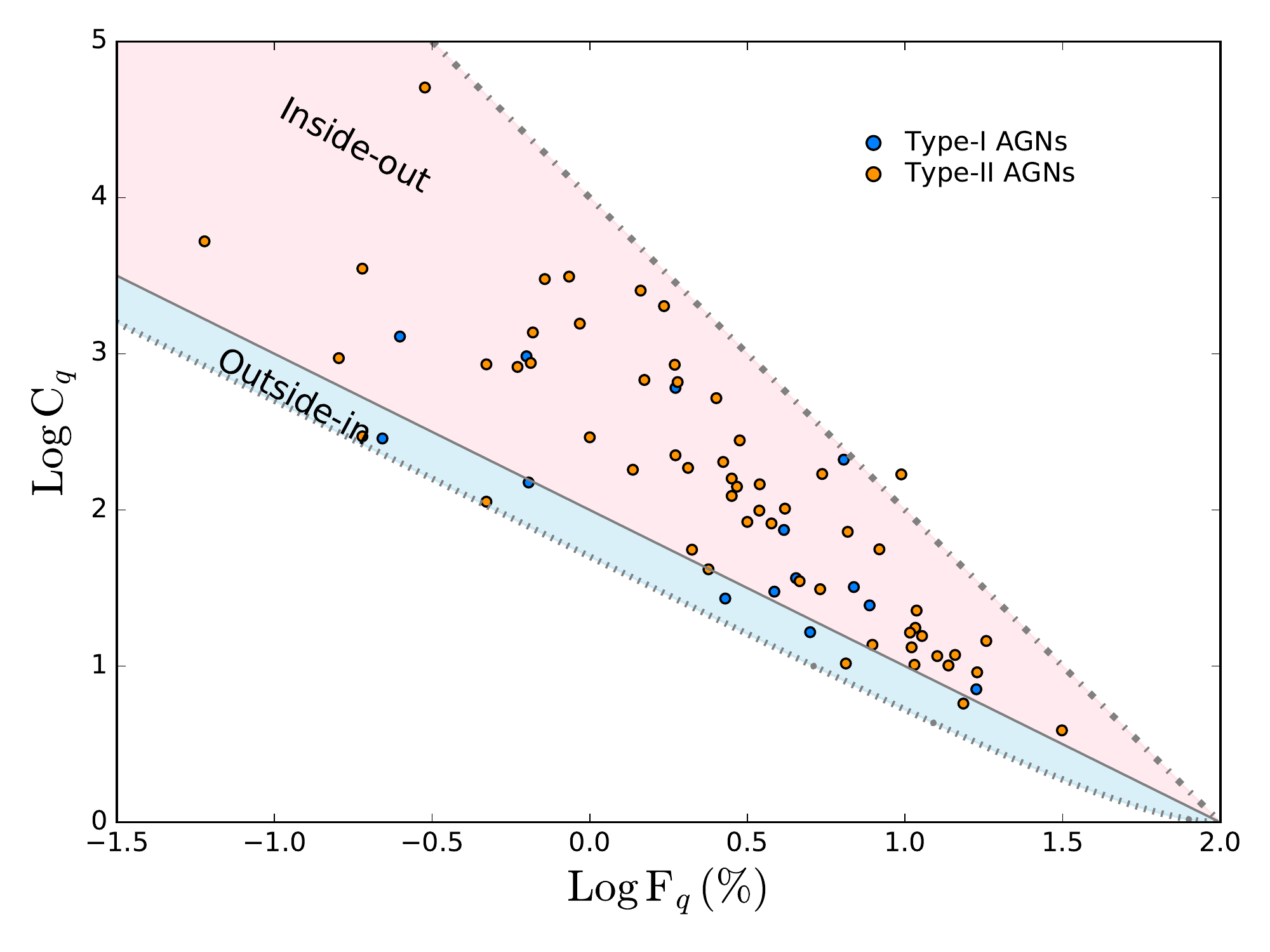}
\caption{The distributions of MaNGA-selected AGN hosts on the quiescence (\fq) vs. quenching concentration (\cq) plane. The Type-I and tyoe-II AGNs are shown in blue and orange symbols, respectively. Similar to figure \ref{fig:fq-cq_model}, the light blue shaded area denotes the outside-in like quenching mode whereas the pink shaded region denotes the inside-out like quenching mode. 
\label{fig:fq-cq_agn}}
\end{figure}

\begin{figure}
\centering
\includegraphics[angle=0,width=0.5\textwidth]{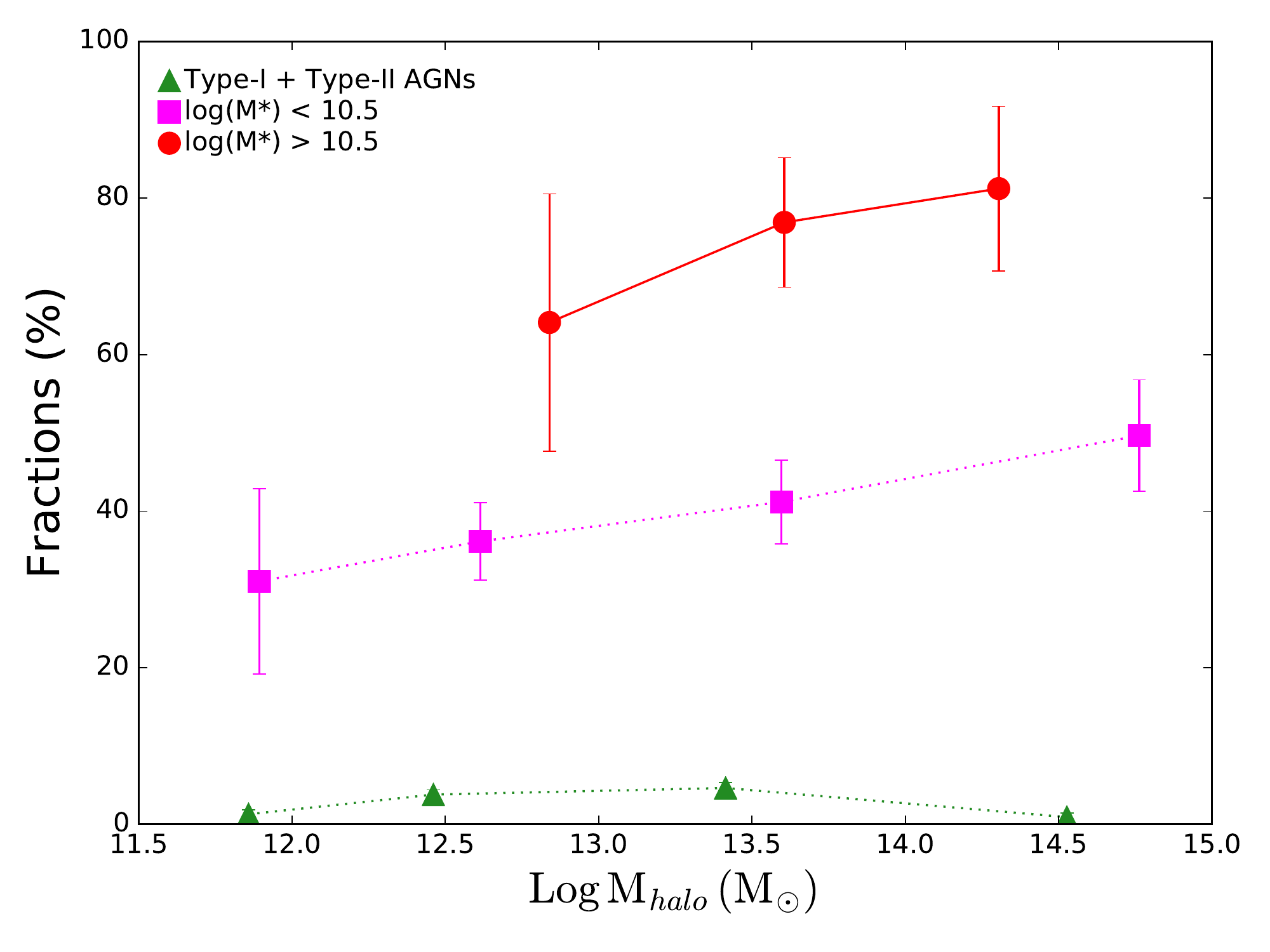}
\caption{The fraction of type-I plus type-II AGNs (green triangles),  high S\'{e}rsic galaxies with \sm$<10^{10.5}$ \Msolar~ (magenta symbols), and high S\'{e}rsic galaxies with \sm$>10^{10.5}$ \Msolar~ (red symbols) as a function of halo mass. The error bars are computed as the root-mean-square normalized by the square root of the sample size in each bin. \label{fig:frac2_halo}}
\end{figure}

\begin{figure}
\centering
\includegraphics[angle=0,width=0.5\textwidth]{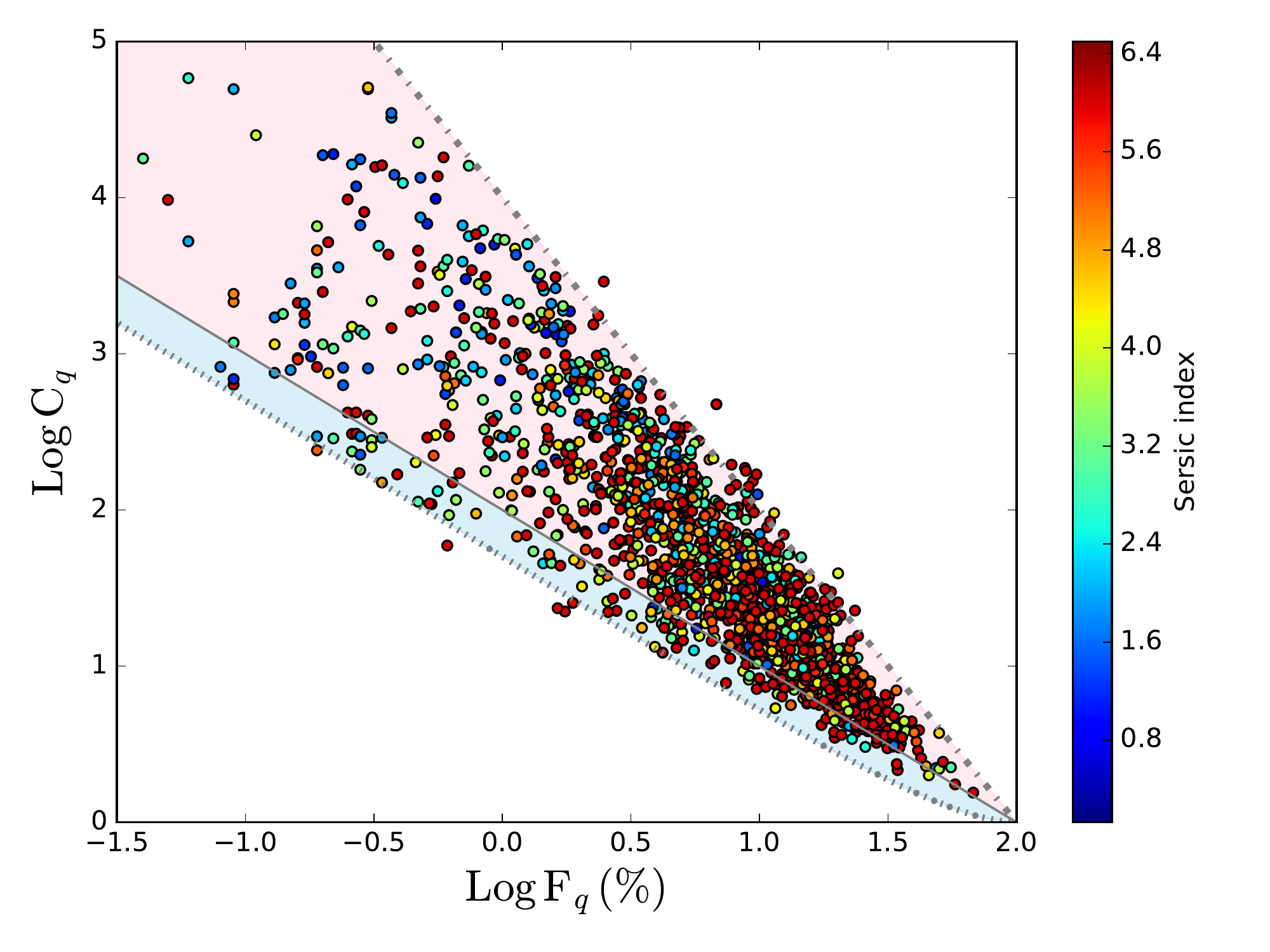}
\caption{The distributions of MaNGA galaxies on the quiescence (\fq) vs. quenching concentration (\cq) plane, color-coded according to the their S\'{e}rsic index. Similar to Figure \ref{fig:fq-cq_model}, the light blue shaded area denotes the outside-in like quenching mode whereas the pink shaded region denotes the inside-out like quenching mode. 
\label{fig:fq-cq_S\'{e}rsic}}
\end{figure}

\begin{deluxetable*}{lcccc}
\tabletypesize{\scriptsize}
\tablewidth{0pt}
\tablecaption{Inside-out and outside-in quchening fractions  as a function of halo mass\label{tab:fq}}
\tablehead{
    \colhead{Subsample} &
    \colhead{Stellar mass cut} &
    \colhead{$M_{halo}$(\Msolar)} &
    \colhead{$f_{in-out}$(\%)} &
    \colhead{$f_{out-in}$(\%)}
}

\startdata
Centrals & \sm$>10^{10.5}$\Msolar & 11.95 & $29.6 \pm 6.9$ & $2.5 \pm 1.8$ \\
 Centrals & \sm$>10^{10.5}$\Msolar & 12.49 & $61.1 \pm 3.6$ & $3.1 \pm 0.6$ \\
 Centrals & \sm$>10^{10.5}$\Msolar & 13.37 & $71.7 \pm 4.7$ & $3.4 \pm 0.8$ \\
 Centrals & \sm$>10^{10.5}$\Msolar & 14.21 & $67.3 \pm 15.2$ & $6.1 \pm 3.6$ \\
 Satellites & $10^{9.5}$\Msolar$<$\sm$<10^{10.5}$\Msolar & 11.88 & $17.2 \pm 8.3$ & $3.4 \pm 3.5$ \\
 Satellites & $10^{9.5}$\Msolar$<$\sm$<10^{10.5}$\Msolar & 12.59 & $29.2 \pm 4.3$ & $4.0 \pm 1.4$ \\
 Satellites & $10^{9.5}$\Msolar$<$\sm$<10^{10.5}$\Msolar & 13.58 & $35.8 \pm 4.9$ & $3.4 \pm 1.3$ \\
 Satellites & $10^{9.5}$\Msolar$<$\sm$<10^{10.5}$\Msolar & 14.67 & $55.8 \pm 7.7$ & $5.4 \pm 2.0$ \\
 Satellites & \sm$>10^{10.5}$\Msolar & 12.83 & $59.0 \pm 15.5$ & $5.1 \pm 3.7$ \\
 Satellites & \sm$>10^{10.5}$\Msolar & 13.60 & $62.3 \pm 7.1$ & $7.0 \pm 1.9$ \\
 Satellites & \sm$>10^{10.5}$\Msolar & 14.30 & $74.4 \pm 9.9$ & $5.3 \pm 2.0$
\enddata

\end{deluxetable*}

\begin{deluxetable}{lccc}
\tabletypesize{\scriptsize}
\tablewidth{0pt}
\tablecaption{Fractions of AGN and galaxies with high S\'{e}rsic indice as a function of halo mass\label{tab:fq_agn_sersic}}
\tablehead{
    \colhead{Subsample} &
    \colhead{Stellar mass cut} &
    \colhead{$M_{halo}$(\Msolar)} &
    \colhead{Fraction(\%)}
}

\startdata

AGN & \sm$>10^{9.5}$\Msolar & 11.86 & $1.3 \pm 0.6$ \\
 AGN & \sm$>10^{9.5}$\Msolar & 12.46 & $3.8 \pm 0.6$ \\
 AGN & \sm$>10^{9.5}$\Msolar & 13.41 & $4.6 \pm 0.7$ \\
 AGN & \sm$>10^{9.5}$\Msolar & 14.53 & $0.9 \pm 0.5$ \\
 $n\geq3$ & $10^{9.5}$\Msolar$<$\sm$<10^{10.5}$\Msolar & 11.89 & $31.0 \pm 11.8$ \\
 $n\geq3$ & $10^{9.5}$\Msolar$<$\sm$<10^{10.5}$\Msolar & 12.61 & $36.1 \pm 4.9$ \\
 $n\geq3$ & $10^{9.5}$\Msolar$<$\sm$<10^{10.5}$\Msolar & 13.60 & $41.2 \pm 5.3$ \\
 $n\geq3$ & $10^{9.5}$\Msolar$<$\sm$<10^{10.5}$\Msolar & 14.76 & $49.7 \pm 7.1$ \\
 $n\geq3$ & \sm$>10^{10.5}$\Msolar & 12.84 & $64.1 \pm 16.4$ \\
 $n\geq3$ & \sm$>10^{10.5}$\Msolar & 13.61 & $76.9 \pm 8.3$ \\
 $n\geq3$ & \sm$>10^{10.5}$\Msolar & 14.30 & $81.2 \pm 10.5$ 

\enddata
\end{deluxetable}

\subsection{The dependence of quiescence (\fq) and concentration (\cq) on the global sSFR}
Our working assumption is that the regions showing LI(N)ER-like emissions and with low EW(\ha) are essentially places where the star formation has already ceased. If this is true, one would expect that more active star-forming galaxies should have lower \fq~whereas quiescent galaxies tend to have high \fq. This correlation is illustrated in Figure \ref{fig:fq-ssfr}, where we plot the quiescence (\fq) vs. the global sSFR computed from Pipe3D.  As expected, the quiescence parameter decreases with increasing global sSFR, although with large scatter. The scatter is greater for galaxies with high sSFR. This is because the strength of the sSFR may differ in HII spaxels among galaxies even at a given fixed fraction of LI(N)ER spaxels. Nevertheless, it is still encouraging to see the  anticorrelation between the defined quiescence and global sSFR. 

Next, we plot the quenching concentration (\cq) as a function of global sSFR in Figure \ref{fig:cq-ssfr}. We find that there is also a fairly good correlation between \cq~and global sSFR. Galaxies with lower sSFR tend to have more extended distributions of quenched area, and hence low \cq. However, we also notice that \cq~  seems to decline with sSFR for galaxies with log(sSFR) $>-10.5$. This is likely due to the small number statistics of both the sample size of galaxies in the two highest log(sSFR) bins and the small number of retired spaxels associated with them.

\subsection{Quiescence (\fq) vs. quenching concentration (\cq)}
In figure \ref{fig:fq-cq_ssfr} we plot the distributions of all MaNGA galaxies on the \fq~versus \cq~ plane, color-coded according to their global sSFR. The two toy model lines are also shown to guide the eyes.  
It can be seen that our sample spreads over the regions between the two model lines on the \fq~versus \cq~ plane. This could be attributed to two effects. First, the quenching may not occur subsequently with increasing or decreasing radius and/or that there could be mixed modes of quenching for most of the galaxies. For example, when a centrally quenched galaxy due to prior AGN or morphological quenching falls into a cluster environments and hence suffers from ram pressure stripping, it can exhibit both the inside-out and outside-in quenching features. Secondly, as illustrated in figures \ref{fig:img_inout} and \ref{fig:img_outin}, the spatial distribution of quenched areas may be patchy, not necessarily axis-symmetric. 

In order to properly classify galaxies into the outside-in and inside-out quenching categories, we analytically compute the \fq~ and \cq~ values by varying the relative contributions from the two modes and investigate how these two parameters change on the \fq~versus \cq~ plane. The detailed calculations are given in the Appendix A. As illustrated in Figure \ref{fig:fq-cq_model_2}, even a small contribution from the outside-in quenching will move the locations toward the pure outside-in quenching line. Therefore, we define a dividing line (gray line), which corresponds to the 50\% inside-out and 50\% outside-in quenching contributions to separate the two types of quenching modes. Galaxies lying below (above) this threshold line are counted as outside-in (inside-out) dominant quenching.

Figure \ref{fig:sm-sfr_quench} presents the global SFR --\sm~distribution derived from the Pipe3D analysis of the full MaNGA sample, color-coded according to their quenching types (red: inside-out; blue: outside-in; grey: unclassified). We split the sample into two stellar mass bins. For massive galaxies (\sm$>10^{10.5}$\Msolar), those objects showing either inside-out (68\%) or outside-in (5\%) quenching features are located in the lower side of the star-forming sequence and the quiescent population. On the other hand, in the low mass bin ($10^{9.5}$\Msolar$<$ \sm $<10^{10.5}$\Msolar), 31\% are classified as inside-out quenching while 5\% are classified as outside-in quenching. These low-mass galaxies with quenching features predominately lie in the quiescent (passive) population. 

\subsection{The halo mass dependence of the inside-out and outside-in quenching}

To see how environments might affect the quenching patterns of galaxies, we make a similar plot by binning the galaxies based on their hosting halo masses for all galaxies with \sm~$> 10^{9.5}$\Msolar~ (top panels of Figure \ref{fig:fq-cq_halo}). For this purpose, we limit our sample to 2915 galaxies that have the halo mass meausurements. As the strength of ram-pressure stripping is proportional to the density of intergalactic medium \citep{gun72}, it has been suggested to take place in denser environments, such as galaxy clusters.  If ram pressure stripping is indeed a dominant process that quenches the star formation in cluster-like halos, we would expect to see fractionally more galaxies classified as outside-in quenching with increasing halo mass. As mentioned in Sec. 4.2, the fractions of galaxies that exhibit inside-out or outside-in quenching features are stellar mass dependent.  In order to remove the stellar mass effect, we also present the results in two stellar mass bins,  $10^{9.5}$\Msolar $<$ \sm$<10^{10.5}$\Msolar~ and \sm~$>10^{10.5}$\Msolar, in the middle and bottom panels, respectively. The two stellar mass cuts adopted here yield a similar dynamical range in terms of stellar mass and are also able to provide sufficient number of satellites in different halo mass bins for subsequent analyses. For centrals and satellites, we compute the fraction of galaxies exhibiting inside-out quenching patterns (\finout) as the following:

\begin{align}\label{eq:finout}
f_{in-out}^{cen.} &= N_{in-out}^{cen.}/N^{cen.}\\
f_{in-out}^{sat.} &= N_{in-out}^{sat.}/N^{sat.},
\end{align}

where $N^{cen.}$ ($N^{sat.}$) is the number of total central (satellite) galaxies and $N_{in-out}^{cen.}$ ($N_{in-out}^{sat.}$) is the number of central (satellite) galaxies classified as inside-out quenching. There are galaxies, however, which do not have LI(N)ER spaxels within 1.5 \Re, meaning that neither their \fq~nor \cq~is computed. These galaxies are typically star-forming galaxies as illustrated in Figure \ref{fig:sm-sfr_quench} and can not be categorized in either inside-out quenching or outside-in quenching. In others words, $N_{in-out}$ + $N_{out-in} \leq$ 1. Therefore, we separately compute the fraction of galaxies exhibiting outside-in quenching patterns (\foutin):

\begin{align}\label{eq: foutin}
f_{out-in}^{cen.} &= N_{out-in}^{cen.}/N^{cen.}\\
f_{out-in}^{sat.} &= N_{out-in}^{sat.}/N^{sat.},
\end{align}

where $N_{out-in}^{cen.}$ ($N_{out-in}^{sat.}$) is the number of central (satellite) galaxies classified as outside-in quenching. The derived values of \finout~ and \foutin~are shown in the upper-right and lower-left corners of each panel, respectively, as well as in Table \ref{tab:fq}.

In the left panel of Figure \ref{fig:frac-halo}, we show the fraction of galaxies exhibiting inside-out quenching patterns (\finout) as a function of halo mass. For centrals, we only present results with \sm~$>10^{10.5}$\Msolar~ as the statistics for less massive centrals are relatively poor in massive halos. We see that \finout~ increases with hosting halo mass for both centrals and satellite galaxies. For the satellites,  \finout~ is significantly higher for high stellar mass galaxies than the low stellar mass ones at a given halo mass. This phenomenon is consistent with the finding that the fraction of centrally suppressed galaxies or central LI(N)ER galaxies increases with the stellar mass \citep{bel17,spi18}. Recalling that not all galaxies possess enough LI(N)ER features to be classified as inside-out quenching or outside-in quenching, \finout~ and \foutin~ does not necessarily sum to one. To see the trend for the fraction of galaxies exhibiting outside-in quenching patterns (\foutin), we plot \foutin~versus halo mass in the right panel of Figure \ref{fig:frac-halo}. 

Interestingly, it is shown that the fraction of outside-in quenching galaxies does not strongly depend on the halo mass, for either centrals or satellites. 
The flat dependence on the halo mass for satellites suggests that the ram-pressure stripping may not be a dominant, at least not the only channel, to suppress the star formation of massive satellites in groups and clusters. 
However, we note that the dividing line (50\% inside-out and 50\% outside-in) adopted here is closer to the pure outside-in trajectory (dotted line) and hence it is possible that our results are affected by the small statistics of the outside-in galaxies. To test the robustness of our results, we also repeat the analyses by using the other two dividing lines that correspond to the 60\% inside-out vs 40\% outside-in and 70\% inside-out vs 30\% outside-in contributions (see Appendix A). We find that the flat dependence of the outside-in fraction on the halo mass still holds in these two cases, despite that the fraction of outside-in quenching galaxies increases as a result while moving the dividing line away from the pure outside-in trajectory,

\section{DISCUSSION}
\subsection{What drives the inside-out quenching?}
Our results show that the fraction of galaxies showing an inside-out quenching pattern, strongly depends on the halo mass, regardless of being a central galaxy or a satellite. The increasing frequency of inside-out quenching with halo mass still holds even we split the satellites into two stellar mass bins. Furthermore, we find that more massive galaxies tend to have higher fractions of inside-out quenching than less massive ones, irrespective of their environments. This is in line with the finding in the literature that high-mass galaxies tend to exhibit suppressed sSFR in the galactic cores as opposed to low-mass galaxies \citep{bel18,liu18,san18}.  Our results suggest that the effect of inside-out quenching depends on both stellar mass and halo mass.

Among  all the mechanisms that may suppress the star formation, 
AGN feedback \citep[e.g.,][]{bow06,cro06,fab12} or morphological quenching \citep{mar09}, may potentially drive the features of the inside-out quenching. The study of molecular gas properties of a preliminary set of MaNGA-selected green valley galaxies suggests that the gas is depleted in an inside-out fashion, possibly attributed to the AGN feedback \citep{lin17b}. In order to investigate whether these processes are responsible for  the halo mass dependence of the inside-out quenching, we first investigate the frequency of AGN and galaxies with high S\'{e}rsic index as a function of halo mass in our sample. For the study of AGNs, we utilize the emission-line selected AGN candidates identified by \citep{san18}, updated with the MPL-6 version. The AGN sample contains both Type-I and Type-II AGNs. It is found that the properties of their host galaxies differ slightly in the SFR vs. stellar mass plane -- Type-I AGN hosts span a wide range of sSFR, from star-forming to quiescent, whereas Type-II AGN hosts are preferentially located in the green valley and the high-sSFR end of the quiescent population \citep[see Figure 4 of ][for details]{san18}. Figure \ref{fig:fq-cq_agn} shows the distributions of two types of AGNs on the \fq~ vs. \cq~plane. It is clear that Type-I and Type-II AGNs are distributed differently in this diagram-- 38\% and 15\% of galaxies hosting a Type-I AGN show the inside-out and outside-in quenching patterns, respectively, compared to Type-II AGNs (66\% vs. 8\%). 
 However, the location of Type-I AGN could be strongly affected by the lack of LI(N)ER detection in the central regions that are heavy ionized by the AGN itself, leading to the deficit of inside-out quenching Type-I AGN seen in the analysis.

In Figure \ref{fig:frac2_halo} and Table \ref{tab:fq_agn_sersic} we show the ratio of the number of Type-I plus Type-II AGNs to the number of total galaxies with \sm~$>10^{9.5}$\Msolar~ as a function of halo mass. It can be seen that the AGN fraction peaks in halos with masses between $10^{12.5}$ to $10^{13.5}$ \Msolar, roughly corresponding to the group scales, and drops towards massive cluster-scale halos. This trend remains similar if we restrict the sample to Type-I or Type-II AGNs only, and is different from the increasing inside-out fraction with halo mass. Our finding that the frequency of optically-selected AGNs peaks on group scales is similar to the trend found for samples of X-ray and radio selected AGNs \citep{san02,bes04,arn09,dav17}.

Figure \ref{fig:fq-cq_S\'{e}rsic} shows how the galaxies with different S\'{e}rsic indices ($n$) are distributed in the \fq~and \cq~ plane. We see that galaxies with higher S\'{e}rsic index tend to occupy the inside-out regions. 70\% and 6\% of galaxies with  $n >$ 3 are classified as inside-out and outside-in quenching, respectively. On the other hand, only 17\% and 1\% of galaxies with  $n <$ 3 are classified as inside-out and outside-in quenching, respectively. The high inside-out quenching fraction seen in galaxies with high $n$  is somewhat expected since it is well-known that galaxies with high S\'{e}rsic index or bulge-to-total (B/T) have lower sSFR \citep{whi15,pan18} and that the bulges tend to have old stellar populations \citep{gon15,mc15,lop18} and lower sSFR \citep{pan18}. In Figure \ref{fig:frac2_halo} (also see Table \ref{tab:fq_agn_sersic}) we show the fraction of satellite galaxies with  $n >$ 3 as a function of halo mass for two stellar mass bins (red symbols). In the case of low-mass satellites for which we have enough statistics, it is shown that the fraction of high-S\'{e}rsic index galaxies strongly increases with halo mass, as seen in the fraction of inside-out galaxies. The trend remains similar if we change the threshold of the S\'{e}rsic index to a higher value of 3.5 or a lower value of 2.5.

Unlike the AGN fraction, which is much smaller compared to the inside-out fraction, the fraction of galaxies with high-S\'{e}rsic index is comparable to that of the inside-out galaxies. The good agreement in the halo mass dependence between the high S\'{e}rsic fraction and inside-out fraction indicates that morphological quenching could be responsible for the growing inside-out quenched galaxies found in more massive halos.  Our results seem to favor morphological quenching \citep{mar09} over the AGN feedback as the primary cause for the similar environment dependence of the inside-out quenching. However, it should be noted that the duty cycles of AGN and the timescale of morphological quenching may be quite different. As discussed in \citet{san18}, the timescale of active AGN phase may be on the order of $\sim 0.1$ Gyr \citep{par07,shu15}, which could be shorter than the star formation quenching timescale itself. On the other hand, morphological quenching operates over longer period once the bulge component forms. Galaxies showing inside-out quenching features may be a result of accumulated quenching events over time in the past rather than an on-going event. Therefore, we can not rule out the possibility that galaxies are quenched when they are located in groups (the 'pre-processing' effect), where the AGN is more frequent as seen in our data, after which they fall into bigger clusters. 

An alternative process that is commonly thought to operate in dense environments is called `strangulation' (or `starvation'), refers to the situation where galaxies fail to replenish  the gas due to the removal of extended gas halo. This effect is suggested to be stronger when galaxies fall into dense environments such as groups or cluster of galaxies \citep{lar80,kaw08}. Strangulation is predicted to suppress the star formation uniformly over the entire galaxy and therefore produces spatial distribution of star formation that is distinct from other mechanisms. It has been argued that strangulation is the primary channel to quench the star formation of satellites \citep{van08,pen15,spi18}, although the exact dependence of environments has not been well constrained. There has been an increasing number of studies finding the global suppression of star formation in green valley galaxies or satellite galaxies by comparing their star formation rate or specific star formation rate gradients with respect to the reference sample \citep{bel18,spi18}, favouring the strangulation quenching scenario. However, it should be stressed that spatially uniform suppression of star formation is not necessarily in contradiction with the inside-out or outside-in quenching defined in this work, which are characterized based on the already quenched areas within the galaxies. As the areas with sSFR below a certain threshold will become a LI(N)ER and hence are defined as quenched regions first, depending on the initial slope of the star formation rate profile, the global reduction of star formation may result in the inside-out or outside-in quenching features. 

\subsection{What drives the outside-in quenching?}

A few processes associated with environments may potentially be responsible for the outside-in quenching features. Ram-pressure stripping has long been suggested to be one of the primary mechanisms that suppress the star formation of galaxies in groups and clusters. When galaxies fall into massive groups or clusters, they experience winds  caused by the its relative motion to the hot intraclustser medium (ICM). As a result, the diffuse interstellar medium (IGM) in the outer parts of galaxies can be stripped, leading to the cessation of star formation. An alternative explanation for the outside-in quenching is through galaxy mergers. Both observations and simulations have demonstrated that  galaxy interactions can induce gas inflow to the central parts of galaxies, trigger starbursts (Barnes \& Hernquist 1996; Cox et al. 2006; Lin et al. 2007; but see Bergvall 2003; Barrera-Ballesteros et al. 2015; Fensch et al. 2017 for different findings) and possibly AGN activities \citep{di05,hop06,ell08}. Recent merger simulations further suggest that the star formation could be suppressed at large  galactocentric radii during galaxy-galaxy interactions as a result of gas inflow while the star formation rate in the central part of galaxies is strongly enhanced, followed by an immediate truncation of star formation after the gas fuel is fully consumed. \citep{mor15}.  Under this circumstance, quenching may possibly occur in an outside-in fashion during some phases of galaxy mergers. Observations and simulations found that galaxy mergers are typically more frequent in dense environments \citep{lin10} and peaks in halos with halo mass  $\sim 10^{13}$ \Msolar~ \citep{jia12}. Therefore, if ram-pressure stripping process  is effective in removing the cold gas from the outskirts of galaxies,  we expect to see higher fraction of galaxies with outside-in patterns in massive halos, especially in clusters of galaxies. On the other hand, if galaxy-galaxy interactions dominate the outside-in quenching, the fraction of outside-in quenching galaxies is expected to peak around the group scales.

As shown in Figure \ref{fig:frac-halo}, the dependence of \foutin~on halo mass is nearly flat for both low mass and high mass satellites, which indicates that neither ram-pressure stripping nor galaxy interactions play a dominant role in producing the outside-in quenching processes. Either both processes may contribute to some degrees or there could be some other mechanism that causes the outside-in quenching features.
Nevertheless, it is worth noting that majority of (partially) quenched satellite galaxies is instead in the inside-out quenching category, which strongly depends on halo mass. In other words, while ram-pressure stripping or galaxy interactions may act in massive halos, they may not be the primary mechanisms that produces a higher fraction of passive galaxies seen in the groups and clusters relative to the field galaxies \citep{lin14,jia17,jia18}.

\subsection{Comparisons with other works}

Since the advent of large IFU surveys, there has been a growing number of environmental studies based on the spatial distributions of stellar populations and star formation rate \citep[e.g.,][]{god17,zhe17,sch17,spi18,med18}. While most studies are based on the gradients of age, star formation rate, or specific star formation rate, our approach probes the spatial distribution of quenched areas, which provides complementary information. Therefore, it would be intriguing to compare our results with previous environmental works based on the IFU observations.

 \citet{spi18} studied the environmental effects by exploring the sSFR gradients in central and satellite galaxies using 1494 MaNGA galaxies. In their study, they classified their galaxies into centrally unsuppressed and suppressed types. The former class on average shows flat profiles in sSFR while the latter exhibits positive slopes in the sSFR gradient. By comparing the sSFR of the satellites relative to the centrals with the same stellar mass, they found a global suppression in the satellites and concluded that their results favor the strangulation scenario. Supposedly if the reduction in the sSFR is spatially uniform,  the centrally suppressed galaxies will display the inside-out quenching feature according to our definition, as the central spaxels become LI(N)ER first. On the other hand, the centrally unsuppressed galaxies will not appear as inside-out nor outside-in quenching since the local spaxels become the LI(N)ER regions simultaneously once their sSFRs drop below the HII threshold. Although it is not straightforward to directly translate our results to the gradient studies given the different natures between our approaches, both works seem to point out that outside-in quenching plays a less vital role in environmental quenching.

On the other hand, the greater fraction of inside-out quenching over outside-in quenching found in our work seems to be in direct conflict with the work carried out by \citet{sch17}, who studied the star formation rate gradients in the SAMI sample and concluded that in dense environments the star formation quenches outside-in.  However, there is a fundamental difference between our analyses--the environments are traced by the local nearest neighbour density in \citet{sch17} whereas in this work we focus on the halo mass. These two environment proxies do not necessarily have one-to-one correspondence. As shown in Figure 12 of \citet{lin16}, there is a wide spread of local density at a given halo mass--at low redshifts the highest density is actually dominated by groups rather than clusters. Thus, the relative strength of various environment effects may differ between our samples. 

\section{CONCLUSIONS}\label{sec:conclusion}

Using the spatially resolved data cubes of 2915 galaxies drawn from SDSS-IV MaNGA, we study the quenching properties of galaxies as a function of halo mass in order to probe environmental quenching effects. We use LI(N)ER regions with low EW(\ha) to trace the quenched areas and we define two non-parametric parameters,  quiescence (\fq) and its concentration (\cq), to quantify the strength and the spatial distribution of the quenched areas. With the combination of these two parameters, we are able to classify galaxies into two categories: inside-out and outside-in quenching, and to study their frequency in different mass of halos. Our results can be summarized as follows:

1. The fraction of galaxies showing inside-out quenching increases with both stellar mass (at a fixed halo mass) and halo mass (at a fixed stellar mass). On the other hands, the frequency of outside-in quenching is almost independent of the halo mass. In nearly all local environments, the frequency of inside-out quenching is higher than the frequency of outside-in quenching at a fixed stellar mass and halo mass. The difference between the two quenching modes is more pronounced for galaxies located in the more massive halos. Our results suggest that inside-out quenching is the dominant quenching mode in all environments.

2. We find that the increasing fraction of galaxies exhibiting a high S\'{e}rsic index (and hence a greater bulge component) with halo mass  is similar to the halo mass dependence of the inside-out quenching, suggesting a plausible link between these two phenomena. On the other hand, the frequency of AGNs peaks at group scales, different from the rising curve of the inside-out quenching with respect to the halo mass. Our result seems to favour the morphological quenching over AGN feedback as a primary mechanism driving the environmental dependence of the inside-out quenching, although this could be affected by the issue of uncertainties in the AGN duty cycle.

 3. The lack of the halo mass dependence of outside-in quenching suggests that neither the ram-pressure stripping, nor the galaxy-galaxy merger, is the dominant process in act in massive halos.  It is likely both mechanisms contribute to the outside-in quenching seen in different environments.
 
 Our method that characterizes the quenched areas provides a complementary approach in investigating the quenching mechanisms with respect to other studies that are based on the spatial gradients in the (specific) star formation rate and stellar populations. This study has revealed that both inside-out and outside-in quenching coexist in different environments and that inside-out quenching dominates in the more massive halos. Combining MaNGA data with future spatially resolved molecular gas observations from ALMA will be key to  further understand the cause of quenching in different environments.

\acknowledgments

We thank the anonymous referee for useful suggestions which improve the clarity of this paper. The work is supported by the Academia Sinica under
the Career Development Award CDA-107-M03 and the Ministry of Science \& Technology of Taiwan
under the grant MOST 107-2119-M-001-024 -. L. Lin acknowledges H. Yee for his useful suggestions to this work. SFS is grateful for the support of a CONACYT (Mexico) grant CB-285080, and funding from the PAPIIT-DGAPA-IA101217(UNAM).
MAF is grateful for financial support from the CONICYT Astronomy Program CAS-CONICYT project No.\,CAS17002, sponsored by the Chinese Academy of Sciences (CAS), through a grant to the CAS South America Center for Astronomy (CASSACA) in Santiago, Chile. IL acknowledges partial financial support from PROYECTO FONDECYT REGULAR 1150345. RR thanks to FAPERGS and CNPq for financial support.

Funding for the Sloan Digital Sky Survey IV has been
provided by the Alfred P. Sloan Foundation, the U.S.
Department of Energy Office of Science, and the Participating Institutions. SDSS-IV acknowledges support
and resources from the Center for High-Performance
Computing at the University of Utah. The SDSS web
site is www.sdss.org. SDSS-IV is managed by the Astrophysical Research Consortium for the Participating
Institutions of the SDSS Collaboration including the
Brazilian Participation Group, the Carnegie Institution
for Science, Carnegie Mellon University, the Chilean
Participation Group, the French Participation Group,
Harvard-Smithsonian Center for Astrophysics, Instituto
de Astrof\'isica de Canarias, The Johns Hopkins University, Kavli Institute for the Physics and Mathematics of the Universe (IPMU) / University of Tokyo, Lawrence
Berkeley National Laboratory, Leibniz Institut f\"ur Astrophysik Potsdam (AIP), Max-Planck-Institut f\"ur Astronomie (MPIA Heidelberg), Max-Planck-Institut f\"ur
Astrophysik (MPA Garching), Max-Planck-Institut f\"ur
Extraterrestrische Physik (MPE), National Astronomical Observatory of China, New Mexico State University,
New York University, University of Notre Dame, Observat\'ario Nacional / MCTI, The Ohio State University,
Pennsylvania State University, Shanghai Astronomical
Observatory, United Kingdom Participation Group, Universidad Nacional Aut\'onoma de M\'exico, University of
Arizona, University of Colorado Boulder, University of
Oxford, University of Portsmouth, University of Utah,
University of Virginia, University of Washington, University of Wisconsin, Vanderbilt University, and Yale University.

\appendix
\section{A. Toy models}

\numberwithin{equation}{section}
\renewcommand{\thefigure}{A\arabic{figure}}
\setcounter{figure}{0}

Let's consider a perfectly circular galaxy with external radius $R$. The inside-out quenching region is a circular section
of the galaxy with a radius $R_1$ around the nucleus. The outside-in quenching region extends from radius $R_2$ to $R$.
We will define $R\equiv 1$ so that $R_1$ and $R_2$ are numerically equivalent to fractions of the total galaxy radius. In that case,
the total circular area of a galaxy is numerically identical to $\pi$.

The quenched fraction $F_q$ of a ``mixed'' galaxy (i.e. presenting both inside-out and outside-in quenching) is given by the fraction
of quenched pixels to the total galaxy area, i.e.

\begin{eqnarray}
F_q&=&\frac{\mbox{internal quenching}+\mbox{external quenching}}{\pi}\\
&=&\frac{\pi R_1^2+(\pi-\pi R_2^2)}{\pi}\\
\label{eq0}
&=&1+R_1^2-R_2^2.
\end{eqnarray}

The contribution of inside-out quenching to the $F_q$ value above, $F_{qI}$, can be expressed as

\begin{eqnarray}
F_{qI}&=&\frac{\pi R_1^2}{\pi R_1^2+(\pi-\pi R_2^2)}\\
&=&\frac{R_1^2}{F_q},
\end{eqnarray}

\noindent so that we can express $F_q$ in terms of $F_{qI}$ as

\begin{equation}
\label{eq1}
F_q=\frac{R_1^2}{F_{qI}}.
\end{equation}

The concentration $C_q$ is defined as

$$C_q=\frac{\sum{r_{all}^2}}{\sum{r_{quenched}^2}}.$$

\noindent Expressing the density of pixels as $\rho$ and considering the infinitesimal limit, the squared summation of radial distances $r$ up to a radius $R$ can be written as

\begin{eqnarray}
\sum r^2&=&\displaystyle\int_0^R\rho \, 2\pi r^3dr\\
&=&\frac{\pi\rho R^4}{2},
\end{eqnarray}

\noindent so that we can express the concentration of a ``mixed'' galaxy as

\begin{eqnarray}
C_q&=&\frac{\pi\rho /2}{\pi\rho R_1^4/2+\pi\rho (1-R_2^4)/2}\\
\label{eq2}
&=&\frac{1}{1+R_1^4-R_2^4},
\end{eqnarray}

\noindent where $\pi\rho /2$ is the result of $\sum{r_{all}^2}$. From equation  \ref{eq1}, 

$$R_1^4=(F_qF_{qI})^2,$$

\noindent and from equation \ref{eq0},

$$R_2^4=(1+R_1^2-F_q)^2,$$

\noindent so that we can re-write the solution for $C_q$ (eq. \ref{eq2}) as

\begin{eqnarray}
C_q&=&\frac{1}{1+(F_qF_{qI})^2-(1+R_1^2-F_q)^2}\\
&=&\frac{1}{1+(F_qF_{qI})^2-(1+F_qF_{qI}-F_q)^2}.
\end{eqnarray}

Now, in order to express $F_q$ and $C_q$ in the same notation as in the paper (log scale, and $F_q$ as a percent) we define

\begin{eqnarray}
x&=&\log_{10} 100\times F_q\\
y&=&\log_{10} C_q,
\end{eqnarray}

\noindent so that the loci of constant $F_{qI}$ values are given by

\begin{equation}\label{eq:math}
y=\log_{10}\frac{1}{1+\left(F_{qI}10^x/100\right)^2-\left(1+F_{qI}10^x/100-10^x/100\right)^2}.
\end{equation}

\noindent Notice that the $F_{qI}$ values are fractions of the quenched area, not percent values.

\begin{figure}
\centering
\includegraphics[angle=0,width=0.5\textwidth]{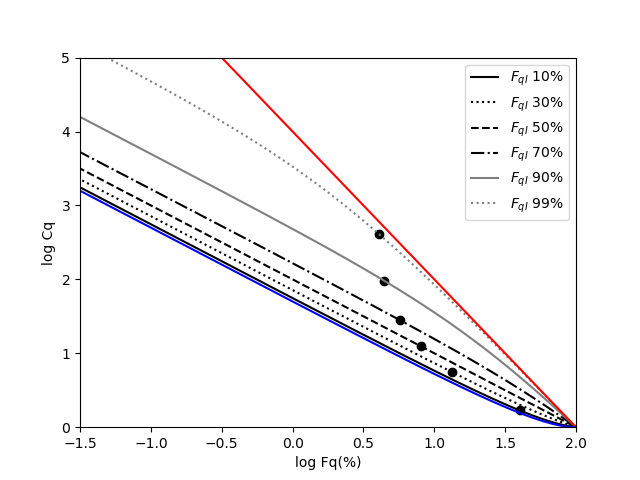}
\caption{The distributions of toy models showing various mixing of inside-out and outside-in quenching on the quiescence (\fq) vs. quenching concentration (\cq) plane using Eq. (A15). The red and blue lines denote the pure inside-out and outside-in quenching lines, respectively. The black lines show the model lines with different fractions of inside-out contributions ($F_{qI}$, increasing from the bottom to the top. The dashed line, corresponding to the 50\%~ inside-out and 50\% outside-in quenching, is used to separate the two modes of quenching in this work. Black circles show the position of a galaxy with inside-out quenching radius equal to 20\% of the total galaxy radius. \label{fig:fq-cq_model_2}}
\end{figure}


\begin{thebibliography}{}
\bibitem[Abdurro'uf(2018)]{abd18} Abdurro'uf, A., Masayuki 2018, \mnras, 479, 5083 
\bibitem[Albareti et al.(2017)]{alb17} Albareti, F.~D., Allende Prieto, C., et al.\ 2017, ApJS accepted, (arXiv:1608.02013)
\bibitem[Argudo-Fern{\'a}ndez et al.(2018)]{arg18} Argudo-Fern{\'a}ndez, M., Lacerna, I., \& Duarte Puertas, S.\ 2018, arXiv:1806.02369 
\bibitem[Arnold et al.(2009)]{arn09} Arnold, T.~J., Martini, P., Mulchaey, J.~S., Berti, A., \& Jeltema, T.~E.\ 2009, \apj, 707, 1691
\bibitem[Baldry et al.(2004)]{bal04} Baldry, I.~K., Glazebrook, K., Brinkmann, J., et al.\ 2004, \apj, 600, 681 
\bibitem[Baldry et al.(2006)]{bal06} Baldry, I.~K., Balogh, M.~L., Bower, R.~G., et al.\ 2006, \mnras, 373, 469
\bibitem[Baldwin et al.(1981)]{bal81} Baldwin, J.~A., Phillips, M.~M., \& Terlevich, R.\ 1981, \pasp, 93, 5 
\bibitem[Barnes \& Hernquist(1996)]{bar96} Barnes, J.~E., \& Hernquist, L.\ 1996, \apj, 471, 115 
\bibitem[Barrera-Ballesteros et al.(2015)]{bar15} Barrera-Ballesteros, J.~K., S{\'a}nchez, S.~F., Garc{\'{\i}}a-Lorenzo, B., et al.\ 2015, \aap, 579, A45
\bibitem[Belfiore et al.(2017)]{bel17} Belfiore, F., Maiolino, R., Tremonti, C., et al.\ 2017, accepted by \mnras~ (arXiv:1703.03813) 
\bibitem[Belfiore et al.(2018)]{bel18} Belfiore, F., Maiolino, R., Bundy, K., et al.\ 2018, \mnras, 477, 3014 
\bibitem[Belfiore et al.(2016)]{bel16} Belfiore, F., Maiolino, R., Maraston, C., et al.\ 2016, \mnras, 461, 3111 
\bibitem[Bergvall et al.(2003)]{ber03} Bergvall, N., Laurikainen, E., \& Aalto, S.\ 2003, \aap, 405, 31 
\bibitem[Best(2004)]{bes04} Best, P.~N.\ 2004, \mnras, 351, 70 
\bibitem[Binette et al.(1994)]{bin94} Binette, L., Magris, C.~G., Stasi{\'n}ska, G., \& Bruzual, A.~G.\ 1994, \aap, 292, 13 
\bibitem[Blanton et al.(2005)]{bla05} Blanton, M.~R., Schlegel, D.~J., Strauss, M.~A., et al.\ 2005, \aj, 129, 2562 
\bibitem[Blanton \& Moustakas(2009)]{bla09} Blanton, M.~R., \& Moustakas, J.\ 2009, \araa, 47, 159 
\bibitem[Blanton et al.(2017)]{bla17} Blanton, M.~R., Bershady, M.~A., Abolfathi, B., et al.\ 2017, \aj submitted (arXiv:1703.00052)
\bibitem[Bower et al.(2006)]{bow06} Bower, R.~G., Benson, A.~J., Malbon, R., et al.\ 2006, \mnras, 370, 645 
\bibitem[Brown et al.(2017)]{bro17} Brown, T., Catinella, B., Cortese, L., et al.\ 2017, \mnras, 466, 1275 
\bibitem[Bryant et al.(2015)]{bry15} Bryant, J.~J., Owers, M.~S., Robotham, A.~S.~G., et al.\ 2015, \mnras, 447, 2857 
\bibitem[Bundy et al.(2015)]{bun15} Bundy, K., Bershady, M. A., Law, D. R., et al. 2015, \apj, 798, 7
\bibitem[Calzetti(2001)]{cal01} Calzetti, D.\ 2001, \pasp, 113, 1449 
\bibitem[Cano-D{\'{\i}}az et al.(2016)]{can16} Cano-D{\'{\i}}az, M., S{\'a}nchez, S.~F., Zibetti, S., et al.\ 2016, \apjl, 821, L26 
\bibitem[Cid Fernandes et al.(2010)]{cid10} Cid Fernandes, R., Stasi{\'n}ska, G., Schlickmann, M.~S., et al.\ 2010, \mnras, 403, 1036 
\bibitem[Cid Fernandes et al.(2011)]{cid11} Cid Fernandes, R., Stasi{\'n}ska, G., Mateus, A., \& Vale Asari, N.\ 2011, \mnras, 413, 1687 
\bibitem[Cid Fernandes et al.(2013)]{cid13} Cid Fernandes, R., P{\'e}rez, E., Garc{\'{\i}}a Benito, R., et al.\ 2013, \aap, 557, A86 
\bibitem[Cooper et al.(2007)]{coo07} Cooper, M.~C., Newman, J.~A., Coil, A.~L., et al.\ 2007, \mnras, 376, 1445 
\bibitem[Cox et al.(2006)]{cox06} Cox, T.~J., Jonsson, P., Primack, J.~R., \& Somerville, R.~S.\ 2006, \mnras, 373, 1013
\bibitem[Croton et al.(2006)]{cro06} Croton, D.~J., Springel, V., White, S.~D.~M., et al.\ 2006, \mnras, 365, 11 
\bibitem[Davies et al.(2017)]{dav17} Davies, R.~I., Hicks, E.~K.~S., Erwin, P., et al.\ 2017, \mnras, 466, 4917
\bibitem[Dressler(1980)]{dre80} Dressler, A.\ 1980, \apj, 236, 351 
\bibitem[Di Matteo et al.(2005)]{di05} Di Matteo, T., Springel, V., \& Hernquist, L.\ 2005, \nat, 433, 604 
\bibitem[Drory et al.(2015)]{dro15} Drory, N., MacDonald, N., Bershady, M.~A., et al.\ 2015, \aj, 149, 77
\bibitem[Elbaz et al.(2011)]{elb11} Elbaz, D., Dickinson, M., Hwang, H.~S., et al.\ 2011, \aap, 533, A119 
\bibitem[Ellison et al.(2008)]{ell08} Ellison, S.~L., Patton, D.~R., Simard, L., \& McConnachie, A.~W.\ 2008, \aj, 135, 1877 
\bibitem[Ellison et al.(2018)]{ell18} Ellison, S.~L., S{\'a}nchez, S.~F., Ibarra-Medel, H., et al.\ 2018, \mnras, 474, 2039 
\bibitem[Fabian(2012)]{fab12} Fabian, A.~C.\ 2012, \araa, 50, 455 
\bibitem[Falc{\'o}n-Barroso et al.(2011)]{fal11} Falc{\'o}n-Barroso, J., S{\'a}nchez-Bl{\'a}zquez, P., Vazdekis, A., et al.\ 2011, \aap, 532, A95
\bibitem[Fensch et al.(2017)]{fen17} Fensch, J., Renaud, F., Bournaud, F., et al.\ 2017, \mnras, 465, 1934
\bibitem[Fischera \& Dopita(2005)]{fis05} Fischera, J., \& Dopita, M.\ 2005, \apj, 619, 340
\bibitem[Fritz et al.(2017)]{fri17} Fritz, J., Moretti, A., Gullieuszik, M., et al.\ 2017, \apj, 848, 132 
\bibitem[Goddard et al.(2017)]{god17} Goddard, D., Thomas, D., Maraston, C., et al.\ 2017, \mnras, 465, 688 
\bibitem[Gomes et al.(2016)]{gom16} Gomes, J.~M., Papaderos, P., Kehrig, C., et al.\ 2016, \aap, 588, A68 
\bibitem[Gonz{\'a}lez Delgado et al.(2014)]{gon14} Gonz{\'a}lez Delgado, R.~M., P{\'e}rez, E., Cid Fernandes, R., et al.\ 2014, \aap, 562, A47 
\bibitem[Gonz{\'a}lez Delgado et al.(2015)]{gon15} Gonz{\'a}lez Delgado, R.~M., Garc{\'{\i}}a-Benito, R., P{\'e}rez, E., et al.\ 2015, \aap, 581, A103
\bibitem[Gonz{\'a}lez Delgado et al.(2016)]{gon16} Gonz{\'a}lez Delgado, R.~M., Cid Fernandes, R., P{\'e}rez, E., et al.\ 2016, \aap, 590, A44 
\bibitem[Gonz{\'a}lez Delgado et al.(2017)]{gon17} Gonz{\'a}lez Delgado, R.~M., P{\'e}rez, E., Cid Fernandes, R., et al.\ 2017, \aap, 607, A128
\bibitem[Gunn \& Gott(1972)]{gun72} Gunn, J. E., \& Gott, J. R. I. 1972, ApJ, 176, 1
\bibitem[Gunn et al.(2006)]{gun06} Gunn, J.~E., Siegmund, W.~A., Mannery, E.~J., et al.\ 2006, \aj, 131, 2332 
\bibitem[Haines et al.(2013)]{hai13} Haines, C.~P., Pereira, M.~J., Smith, G.~P., et al.\ 2013, \apj, 775, 126 
\bibitem[Hopkins et al.(2006)]{hop06} Hopkins, P.~F., Hernquist, L., Cox, T.~J., et al.\ 2006, \apjs, 163, 1
\bibitem[Hsieh et al.(2017)]{hsi17}  Hsieh, B.~C., Lin, L., Lin, J.~H., et al.\ 2017, \apjl, 851, L24 
\bibitem[Huang \& Kauffmann(2015)]{hua15} Huang, M.-L., \& Kauffmann, G.\ 2015, \mnras, 450, 1375 
\bibitem[Ibarra-Medel et al.(2016)]{iba16} Ibarra-Medel, H.~J., S{\'a}nchez, S.~F., Avila-Reese, V., et al.\ 2016, \mnras, 463, 2799 
\bibitem[Jian et al.(2012)]{jia12} Jian, H.-Y., Lin, L., \& Chiueh, T.\ 2012, \apj, 754, 26 
\bibitem[Jian et al.(2017)]{jia17} Jian, H.-Y., Lin, L., Lin, K.-Y., et al.\ 2017, \apj, 845, 74 
\bibitem[Jian et al.(2018)]{jia18} Jian, H.-Y., Lin, L., Oguri, M., et al.\ 2018, \pasj, 70, S23 
\bibitem[Kawata \& Mulchaey(2008)]{kaw08} Kawata, D., \& Mulchaey, J.~S.\ 2008, \apjl, 672, L103 
\bibitem[Kauffmann et al.(2003)]{kau03} Kauffmann, G., Heckman, T. M., White, S. D. M., et al. 2003, \mnras, 341, 33
\bibitem[Kauffmann(1996)]{kau96} Kauffmann, G.\ 1996, \mnras, 281, 487 
\bibitem[Kuntschner et al.(2002)]{kun02} Kuntschner, H., Smith, R.~J., Colless, M., et al.\ 2002, \mnras, 337, 172 
\bibitem[Kewley et al.(2001)]{kew01} Kewley, L.~J., Dopita, M.~A., Sutherland, R. S., Heisler, C.~A., \& Trevena, J.\ 2001, \apj, 556, 121 
\bibitem[Kewley et al.(2006)]{kew06} Kewley, L.~J., Groves, B., Kauffmann, G., \& Heckman, T.\ 2006, \mnras, 372, 961 
\bibitem[Koyama et al.(2013)]{koy13} Koyama, Y., Smail, I., Kurk, J., et al.\ 2013, \mnras, 434, 423 
\bibitem[Lacerna et al.(2016)]{lac16} Lacerna, I., Hern{\'a}ndez-Toledo, H.~M., Avila-Reese, V., Abonza-Sane, J., \& del Olmo, A.\ 2016, \aap, 588, A79 
\bibitem[Lacerna et al.(2018)]{lac18} Lacerna, I., Argudo-Fern{\'a}ndez, M., \& Duarte Puertas, S.\ 2018, arXiv:1806.02368 
\bibitem[Larson et al.(1980)]{lar80} Larson, R. B., Tinsley, B. M., \& Caldwell, C. N. 1980, \apj, 237, 692
\bibitem[Law et al.(2015)]{law15} Law, D.~R., Yan, R., Bershady, M.~A., et al.\ 2015, \aj, 150, 19 
\bibitem[Li et al.(2015)]{li15} Li, C., Wang, E., Lin, L., et al.\ 2015, \apj, 804, 125 
\bibitem[Lin et al.(2007)]{lin07} Lin, L., Koo, D.~C., Weiner, B.~J., et al.\ 2007, \apjl, 660, L51 
\bibitem[Lin et al.(2010)]{lin10} Lin, L., Cooper, M.~C., Jian, H.-Y., et al.\ 2010, \apj, 718, 1158 
\bibitem[Lin et al.(2014)]{lin14} Lin, L., Jian, H.-Y., Foucaud, S., et al.\ 2014, \apj, 782, 33 
\bibitem[Lin et al.(2016)]{lin16} Lin, L., Capak, P.~L., Laigle, C., et al.\ 2016, \apj, 817, 97 
\bibitem[Lin et al.(2017a)]{lin17a} Lin, L., Lin, J.-H., Hsu, C.-H., et al.\ 2017a, \apj, 837, 32 
\bibitem[Lin et al.(2017b)]{lin17b} Lin, L., Belfiore, F., Pan, H.-A., et al.\ 2017b, \apj, 851, 18 
\bibitem[Liu et al.(2018)]{liu18} Liu, F.~S., Jia, M., Yesuf, H.~M., et al.\ 2018, \apj, 860, 60 
\bibitem[L{\'o}pez Fern{\'a}ndez et al.(2018)]{lop18} L{\'o}pez Fern{\'a}ndez, R., Gonz{\'a}lez Delgado, R.~M., P{\'e}rez, E., et al.\ 2018, \aap, 615, A27 
\bibitem[Martig et al.(2009)]{mar09} Martig, M., Bournaud, F., Teyssier, R., \& Dekel, A.\ 2009, \apj, 707, 250 
\bibitem[Martin et al.(2007)]{mar07} Martin, D.~C., Wyder, T.~K., Schiminovich, D., et al.\ 2007, \apjs, 173, 342 
\bibitem[Martins et al.(2005)]{mar05} Martins, L.~P., Gonz{\'a}lez Delgado, R.~M., Leitherer, C., Cervi{\~n}o, M., \& Hauschildt, P.\ 2005, \mnras, 358, 49 
\bibitem[Medling et al.(2018)]{med18} Medling, A.~M., Cortese, L., Croom, S.~M., et al.\ 2018, \mnras, 475, 5194 
\bibitem[McCarthy et al.(2008)]{mc08} McCarthy, I.~G., Frenk, C.~S., Font, A.~S., et al.\ 2008, \mnras, 383, 593 
\bibitem[McDermid et al.(2015)]{mc15} McDermid, R.~M., Alatalo, K., Blitz, L., et al.\ 2015, \mnras, 448, 3484 
\bibitem[Mihos \& Hernquist(1994)]{mih94} Mihos, J. C., \& Hernquist, L. 1994, \apj, 431, L9
\bibitem[Moore et al.(1996)]{moo96} Moore, B., Katz, N., Lake, G., Dressler, A., \& Oemler, A, 1996, Nature, 379, 613
\bibitem[Moreno et al.(2015)]{mor15} Moreno, J., Torrey, P., Ellison, S.~L., et al.\ 2015, \mnras, 448, 1107 
\bibitem[Muzzin et al.(2012)]{muz12} Muzzin, A., Wilson, G., Yee, H.~K.~C., et al.\ 2012, \apj, 746, 188 
\bibitem[Pan et al.(2015)]{pan15} Pan, Z., Li, J., Lin, W., et al.\ 2015, \apjl, 804, L42 
\bibitem[Pan et al.(2018)]{pan18} Pan, H.-A., Lin, L., Hsieh, B.-C., et al.\ 2018, \apj, 854, 159 
\bibitem[Parma et al.(2007)]{par07} Parma, P., Murgia, M., de Ruiter, H.~R., et al.\ 2007, \aap, 470, 875 
\bibitem[Peng et al.(2010)]{pen10} Peng, Y.-j., Lilly, S.~J., Kova{\v c}, K., et al.\ 2010, \apj, 721, 193 
\bibitem[Peng et al.(2015)]{pen15} Peng, Y., Maiolino, R., \& Cochrane, R.\ 2015, \nat, 521, 192 
\bibitem[Papaderos et al.(2013)]{pap13} Papaderos, P., Gomes, J.~M., V{\'{\i}}lchez, J.~M., et al.\ 2013, \aap, 555, L1 
\bibitem[P{\'e}rez et al.(2013)]{per13} P{\'e}rez, E., Cid Fernandes, R., Gonz{\'a}lez Delgado, R.~M., et al.\ 2013, \apjl, 764, L1 
\bibitem[Roberts \& Haynes(1994)]{rob94} Roberts, M.~S., \& Haynes, M.~P.\ 1994, \araa, 32, 115 
\bibitem[S{\'a}nchez \& Gonz{\'a}lez-Serrano(2002)]{san02} S{\'a}nchez, S.~F., \& Gonz{\'a}lez-Serrano, J.~I.\ 2002, \aap, 396, 773 
\bibitem[S{\'a}nchez-Bl{\'a}zquez et al.(2006)]{san06} S{\'a}nchez-Bl{\'a}zquez, P., Peletier, R.~F., Jim{\'e}nez-Vicente, J., et al.\ 2006, \mnras, 371, 703
\bibitem[S{\'a}nchez et al.(2012)]{san12} S{\'a}nchez, S.~F., Kennicutt, R.~C., Gil de Paz, A., et al.\ 2012, \aap, 538, A8 
\bibitem[S{\'a}nchez et al.(2013)]{san13} S{\'a}nchez, S.~F., Rosales-Ortega, F.~F., Jungwiert, B., et al.\ 2013, \aap, 554, A58 
\bibitem[S{\'a}nchez et al.(2015)]{san15} S{\'a}nchez, S.~F., P{\'e}rez, E., Rosales-Ortega, F.~F., et al.\ 2015, \aap, 574, A47 
\bibitem[S{\'a}nchez et al.(2016a)]{san16a} S{\'a}nchez, S. F., P{\'e}rez, E., S{\'a}nchez-Bl{\'a}zquez, P., et al. 2016a, \rmxaa, 52, 21
\bibitem[S{\'a}nchez et al.(2016b)]{san16b} S{\'a}nchez, S. F., P{\'e}rez, E., S{\'a}nchez-Bl{\'a}zquez, P., et al. 2016b, \rmxaa, 52, 171
\bibitem[S{\'a}nchez et al.(2018)]{san18} S{\'a}nchez, S.~F., Avila-Reese, V., Hernandez-Toledo, H., et al.\ 2018, \rmxaa, 54, 217
\bibitem[Sarzi et al.(2010)]{sar10} Sarzi, M., Shields, J.~C., Schawinski, K., et al.\ 2010, \mnras, 402, 2187 
\bibitem[Schaefer et al.(2017)]{sch17} Schaefer, A.~L., Croom, S.~M., Allen, J.~T., et al.\ 2017, \mnras, 464, 121 
\bibitem[Shulevski et al.(2015)]{shu15} Shulevski, A., Morganti, R., Barthel, P.~D., et al.\ 2015, \aap, 583, A89 
\bibitem[Singh et al.(2013)]{sin13} Singh, R., van de Ven, G., Jahnke, K., et al.\ 2013, \aap, 558, A43 
\bibitem[Smee et al.(2013)]{sme13} Smee, S.~A., Gunn, J.~E., Uomoto, A., et al.\ 2013, \aj, 146, 32 
\bibitem[Spindler et al.(2018)]{spi18} Spindler, A., Wake, D., Belfiore, F., et al.\ 2018, \mnras, 476, 580 
\bibitem[Stasi{\'n}ska et al.(2008)]{sta08} Stasi{\'n}ska, G., Vale Asari, N., Cid Fernandes, R., et al.\ 2008, \mnras, 391, L29 
\bibitem[Tacchella et al.(2015)]{tac15} Tacchella, S., Carollo, C.~M., Renzini, A., et al.\ 2015, Science, 348, 314 
\bibitem[Trager et al.(2008)]{tra08} Trager, S.~C., Faber, S.~M., \& Dressler, A.\ 2008, \mnras, 386, 715
\bibitem[van den Bosch et al.(2008)]{van08} van den Bosch, F.~C., Aquino, D., Yang, X., et al.\ 2008, \mnras, 387, 79 
\bibitem[Vazdekis et al.(2010)]{vaz10} Vazdekis, A., S{\'a}nchez-Bl{\'a}zquez, P., Falc{\'o}n-Barroso, J., et al.\ 2010, \mnras, 404, 1639
\bibitem[Veilleux \& Osterbrock(1987)]{vei87} Veilleux, S., \& Osterbrock, D.~E.\ 1987, \apjs, 63, 295 
\bibitem[Vogt et al.(2013)]{vog13} Vogt, F. P. A., Dopita, M. A., \& Kewley, L. J. 2013, \apj, 768, 151 
\bibitem[Vulcani et al.(2010)]{vul10} Vulcani, B., Poggianti, B.~M., Finn, R.~A., et al.\ 2010, \apjl, 710, L1 
\bibitem[Wang et al.(2018)]{wan18} Wang, E., Li, C., Xiao, T., et al.\ 2018, \apj, 856, 137 
\bibitem[Wetzel et al.(2012)]{wet12} Wetzel, A.~R., Tinker, J.~L., \& Conroy, C.\ 2012, \mnras, 424, 232 
\bibitem[Whitaker et al.(2015)]{whi15} Whitaker, K.~E., Franx, M., Bezanson, R., et al.\ 2015, \apjl, 811, L12
\bibitem[Yan \& Blanton(2012)]{yan12} Yan, R., \& Blanton, M.~R.\ 2012, \apj, 747, 61 
\bibitem[Yan et al.(2016a)]{yan16a} Yan, R., Tremonti, C., Bershady, M.~A., et al.\ 2016a, \aj, 151, 8
\bibitem[Yan et al.(2016b)]{yan16b} Yan, R., Bundy, K., Law, D.~R., et al.\ 2016b, \aj, 152, 197
\bibitem[Yang et al.(2007)]{yan07} Yang, X., Mo, H.~J., van den Bosch, F.~C., et al.\ 2007, \apj, 671, 153 
\bibitem[Yang et al.(2008)]{yan08} Yang, X., Mo, H.~J., \& van den Bosch, F.~C.\ 2008, \apj, 676, 248 
\bibitem[Zhang et al.(2017)]{zha17} Zhang, K., Yan, R., Bundy, K., et al.\ 2017, \mnras, 466, 3217 
\bibitem[Zheng et al.(2017)]{zhe17} Zheng, Z., Wang, H., Ge, J., et al.\ 2017, \mnras, 465, 4572 













\end{thebibliography}
\end{document}